\documentclass[12pt,preprint]{aastex}

\slugcomment{Accepted by ApJ November 28 2012}

\shorttitle{Y dwarfs and new cool cloudy models}
\shortauthors{Leggett et al.}

\begin{document}


\title{A Comparison of Near-Infrared
Photometry and Spectra for Y Dwarfs with a New Generation of Cool Cloudy
Models}


\author{S. K. Leggett\altaffilmark{1}}
\email{sleggett@gemini.edu}
\author{Caroline V. Morley\altaffilmark{2}}
\author{M. S. Marley\altaffilmark{3}}
\author{D. Saumon\altaffilmark{4}}
\author{Jonathan J. Fortney\altaffilmark{2}}
\and
\author{Channon Visscher\altaffilmark{5}}

\altaffiltext{1}{Gemini Observatory, Northern Operations Center, 670
  N. A'ohoku Place, Hilo, HI 96720, USA} 
\altaffiltext{2}{Department of Astronomy and Astrophysics, University of California,
Santa Cruz, CA 95064, USA}
\altaffiltext{3}{NASA Ames Research Center, Mail Stop 245-3, Moffett Field, CA 94035, USA}
\altaffiltext{4}{Los Alamos National Laboratory, PO Box 1663, MS F663, Los Alamos, NM 87545, USA}
\altaffiltext{5}{Southwest Research Institute, Boulder, CO 80302, USA}

\begin{abstract}

We present $YJHK$ photometry, or a subset, for the six Y dwarfs discovered in {\it WISE} data by 
Cushing et al.. The data were obtained using NIRI on the Gemini North telescope; $YJHK$ were obtained 
for WISEP J041022.71$+$150248.5, WISEP J173835.52$+$273258.9 and WISEPC J205628.90$+$145953.3; $YJH$ 
for WISEPC J140518.40$+$553421.5 and WISEP J154151.65$–$225025.2; $YJK$ for WISEP 
J182831.08$+$265037.8. We also present a far-red spectrum obtained using GMOS-North for WISEPC 
J205628.90$+$145953.3. We compare the data to Morley et al. (2012) models, which include cloud decks 
of sulfide and chloride condensates. We find that the models with these previously neglected clouds 
can reproduce the energy distributions of T9 to Y0 dwarfs quite well, other than near 5 $\mu$m 
 where the models are too bright. This is thought to be because the models do not include 
departures from chemical equilibrium caused by vertical mixing, which would enhance the abundance of 
CO and CO$_2$, decreasing the flux at 5 $\mu$m. Vertical mixing also decreases the abundance of NH$_3$, 
which would otherwise have strong absorption features at 1.03 $\mu$m and 1.52 $\mu$m that are not seen 
in the Y0 WISEPC J205628.90$+$145953.3.  We find that the 
five Y0 to Y0.5 dwarfs have $300 \lesssim T_{\rm eff}$~K$ \lesssim 450$, $4.0 \lesssim \log g \lesssim 
4.5$ and $f_{\rm sed} \approx 3$. These temperatures and gravities imply a mass range of 5 -- 15 
M$_{\rm Jupiter}$ and ages around 5~Gyr. We suggest that WISEP J182831.08$+$265037.8 is a binary 
system, as this better explains its luminosity and color. We find that the data can be made consistent 
with observed trends, and generally consistent with the models, if the system is composed of a $T_{\rm 
eff} \approx 325$~K and $\log g \lesssim 4.5$ primary, and a $T_{\rm eff} \approx 300$~K and $\log g 
\gtrsim 4.0$ secondary, corresponding to masses of 10 and 7 M$_{\rm Jupiter}$ and an age around 2~Gyr. 
If our deconvolution is correct, then the $T_{\rm eff} \approx 300$~K cloud-free model fluxes at $K$ 
and W2 are too faint by 0.5 -- 1.0 magnitudes. We will address this discrepancy in our next generation 
of models, which will incorporate water clouds and mixing.

\end{abstract}

\keywords{stars: brown dwarfs, Stars: atmospheres}


\section{Introduction}

Over the last 15 years, far-red and near-infrared ground-based sky surveys have revealed the 
local-neighborhood populations of L and T dwarfs (e.g. Kirkpatrick 1995). The 2-m-class telescopes of 
the Two Micron All Sky Survey (2MASS, Skrutskie et al. 2006) and the Sloan Digital Sky Survey (SDSS, 
York et al. 2000) extended the brown dwarf sequence to T8 spectral type with effective temperatures 
($T_{\rm eff}$) as low as 750~K (e.g. Burgasser et al. 2000).  The 4-m-class UKIRT Infrared Deep Sky 
Survey (UKIDSS, Lawrence et al. 2007) and Canada France Brown Dwarf Survey (CFBDS, Delorme et al. 
2008) identified dwarfs as cool as $T_{\rm eff} = 500$~K, with spectral types of T9 or T10 (e.g. Lucas 
et al. 2010). In 2011, the mid-infrared 0.4-m telescope Wide-field Infrared Survey Explorer ({\it 
WISE}, Wright et al. 2010) revealed prototype Y dwarfs (Cushing et al. 2011), with $T_{\rm eff} =$ 300 
-- 450~K.  The extension of the lower main-sequence to the brown dwarfs now spans a range in 
luminosity of $10^{-4}$ to $10^{-7}$ L$_{\odot}$, and in mass of $\sim 70$ to 10 M$_{\rm Jupiter}$.

The T and Y dwarfs are more similar to the gas-giant planets than to the stars, and clouds complicate 
the analysis of their spectra.  The spectral energy distributions (SEDs) of L dwarfs are reddened by 
clouds consisting of liquid and solid iron and silicates (e.g. Ruiz, Leggett \& Allard 1997; Ackerman 
\& Marley 2001; Tsuji 2002; Helling et al. 2008). The transition from L to T is associated with a 
clearing of these clouds, such that mid-type T dwarfs have clear atmospheres (e.g. Knapp et al. 2004; 
Marley, Saumon \& Goldblatt 2010).  We have recently shown (Morley et al. 2012) that chloride and 
sulfide clouds are important in the atmospheres of late-T and early-Y dwarfs. First water and then 
ammonia clouds are expected to occur as dwarfs cool below $T_{\rm eff} \sim 400$~K, as seen in 
Jupiter.

In this work we present new near-infrared $YJHK$ photometry for the six objects identified by Cushing et al. (2011) as Y dwarfs. These data are more precise and cover a wider wavelength range than the $JH$ photometry presented in Cushing et al. and Kirkpatrick et al. (2012). We compare the new and previously published data to the models which include the chloride and sulfide clouds, reconfirm the validity of the models, and use them to estimate temperature and gravity for the Y dwarfs. We also present a far-red spectrum of one of the brightest Y dwarfs, obtained as part of a search for NH$_3$ absorption features in 400~K objects.

\section{Observations}

\subsection{NIRI Photometry}

Photometry was obtained in some or all  of the $YJHK$ filters for the six objects identified by Cushing et al. (2011) as Y dwarfs: WISEP J041022.71$+$150248.5, WISEPC J140518.40$+$553421.5, WISEP J154151.65$–$225025.2, WISEP J173835.52$+$273258.9, WISEP J182831.08$+$265037.8 and WISEPC J205628.90$+$145953.3. Hereafter the source names are abbreviated to the first four digits of the RA and Declination. All of $YJHK$ were measured for the three brightest objects WISEPC J0410$+$1502, WISEPC J1738$+$2732 and WISEPC J2056$+$1459. $YJH$ only were obtained for WISEPC J1405$+$5534 and WISEPC J1541$-$2250 --- due to constraints on available telescope time the fainter $K$-band observations were omitted. WISEPC J1828$+$2650 was identified by Cushing et al. as the coolest of the Y dwarfs, and we obtained $YJK$ photometry for this particularly interesting object; we could not improve on the accuracy of the $H$ magnitude given by Kirkpatrick et al. (2012) in a reasonable amount of time, and so $H$-band data were not obtained. The $YJH$ photometry for  WISEPC J1405$+$5534 and the $YJ$ photometry for   WISEPC J1541$-$2250 was previously published by Morley et al. (2012).

The Near-Infrared Imager (NIRI, Hodapp et al. 2003) was used on Gemini North in programs GN-2012A-DD-7, GN-2012A-Q-106, GN-2012B-Q-27 and GN-2012B-Q-75. The filter sets are on the Mauna Kea Observatories (MKO) system (Tokunaga, Simons \& Vacca 2002, Tokunaga \& Vacca 2005), although there is some variation in the $Y$ filter bandpass between the cameras used on Mauna Kea (see \S 3 and Liu et al. 2012).

Exposure times of 30~s or 60~s were used, with a 5- or 9-position telescope dither pattern. The total integration time is given in Table 1, together with the derived photometry. The data were reduced in the standard way using routines supplied in the IRAF Gemini package. UKIRT Faint Standards  were used for calibration, taking the $Y$ data from the UKIRT online catalog\footnote{http://www.jach.hawaii.edu/UKIRT/astronomy/calib/phot cal/fs ZY MKO wfcam.dat} and the $JHK$ data from Leggett et al. (2006). All data were taken on photometric nights with typical near-infrared seeing of  $0\farcs 8$. Aperture photometry was carried out with apertures of radii 5 -- 8 pixels, or diameters of $1\farcs 2$ -- $1\farcs 9$; 
aperture corrections were derived from stars in the field. Sky levels were determined from concentric annular regions and uncertainties were derived from the sky variance.

Most of our measurements agree within the errors with the MKO-system values presented in Cushing et al. (2011) and Kirkpatrick et al. (2012). However two of the dwarfs are much fainter in $J$ than the previously published values:  WISEPC J1405$+$5534  is a magnitude fainter, and WISEPC J1738$+$2732 is 0.5 magnitudes fainter. Also one dwarf is a magnitude fainter at $H$,  WISEPC J0410$+$1502.
Our measurements are more consistent with the shape of the near-infrared spectra presented by Cushing et al., and it seems likely that the Palomar WIRC data are corrupted, perhaps by detector hot pixels or similar, and that the difference is not due to extreme variability.

\subsection{GMOS Spectroscopy}

We obtained far-red spectra of WISEPC J2056$+$1459 using the Gemini Multi-Object Spectrograph
(GMOS, Hook et al. 2004) at Gemini North, through Director's Discretionary Time granted under program
GN-2012A-DD-7. The R150 grating was used with the RG610 blocking filter.  The central wavelength was 1000~nm, with wavelength coverage of 600 --- 1040 nm, though the Y dwarf is only detected at wavelength longer than 830~nm.        
The 1$\farcs$5 slit was used with $2\times 2$ binning, and the resulting resolution was $R \approx$ 900 or 10 \AA.
Four 3200~s frames were obtained on 2012 June 5, for a total on-source time of 3.6 hours.
Flatfielding and wavelength calibration were achieved using lamps in the on-telescope calibration unit. 
The spectrophotometric standard BD +28 4211 was used to determine the instrument response curve.
The data were reduced using routines  supplied in the IRAF Gemini package. 
The spectrum was flux calibrated using the measured NIRI $Y$ photometry, extending the GMOS spectrum from 1.04~$\mu$m to 1.09~$\mu$m using the Cushing et al. (2011) $Y$-band spectrum of WISEPC J1541$-$2250 as a template. 

Figure 1 shows the GMOS spectrum for WISEPC J2056$+$1459, as well as the spectrum for UGPS J072227.51$-$054031.2 (UGPS J0722$-$0540) from Leggett et al. (2012). The absorption features due to Cs I that are apparent in the $T_{\rm eff} = 500$~K dwarf UGPS J0722$-$0540 are not detected in the $T_{\rm eff} \approx 400$~K dwarf WISEPC J2056$+$1459. This is not unexpected --- Cs should be predominantly in the form of CsCl in atmospheres this cool (e.g. Lodders 1999, Leggett et al. 2012). What is unexpected is the lack of strong NH$_3$ features at 1.02 -- 1.04 $\mu$m (e.g. Leggett et al. 2007b). We return to this point below in \S 4.4.

\section{Photometry and Astrometry}

Kirkpatrick et al. (2011, 2012) give trigonometric parallaxes for WISEPC J0410$+$1502, WISEPC J1405$+$5534,  WISEPC J1541$-$2250,
WISEPC J1738$+$2732 and WISEPC J1828$+$2650. Table 2 lists these values, together with the NIRI photometry presented in \S 2.1, and the $H$ band magnitude for WISEPC J1828$+$2650 presented in Kirkpatrick et al. (2012).  Liu et al. (2012, their Appendix A) show that small differences in the $Y$ bandpass can lead to significant differences in T and Y dwarf photometry. The NIRI $Y$ filter is shifted blueward of the WFCAM/UKIDSS MKO-$Y$ filter by 0.007~$\mu$m or $\sim$7\% of the filter width (the Keck  NIRC2 filter is bluer than the NIRI filter by a similar amount). Due to the rapidly rising flux to the red side of the filter, the blueward shift results in fainter NIRI magnitudes. Liu et al. use spectra of T8 -- Y0 dwarfs to  synthesize photometry in the NIRI and UKIDSS systems, and derive $Y_{NIRI} - Y_{WFCAM} = 0.17 \pm 0.03$. As the largest set of $Y$-band photometry has been obtained via the UKIDSS project, it is assumed that the UKIRT/WFCAM filter defines the $Y$-band of the MKO photometric system, and in the plots shown in this paper we have reduced the NIRI $Y$-band magnitudes by 0.17.

Kirkpatrick et al. also give the All-Sky Data Release {\it WISE} photometry for the six Y dwarfs, as well as {\it Spitzer} warm-mission IRAC 3.6~$\mu$m and 4.5~$\mu$m photometry. IRAC data obtained by the {\it WISE} team is now available in the Spitzer Science Archive. The data were obtained between 2010 July and 2011 March,  via 
cycle 7 GO program 70062 and DDT program 551, with PIs Kirkpatrick and Mainzer.
We have carried out aperture photometry on the archived data, using the mosaics produced by the Spitzer pipeline S18.18.0. For all but one source, apertures with radii of 6 pixels or diameter 7.2" were used; due to a nearby star, smaller apertures were used for WISEPC J1541$-$2250 of radius 3 and 4.5 pixels for the 3.6~$\mu$m and 4.5~$\mu$m channels respectively. Aperture corrections were derived from stars in the field for WISEPC J1541$-$2250, and from the IRAC handbook\footnote {http://ssc.spitzer.caltech.edu/irac/dh/} for the other brown dwarfs. In all cases sky levels were determined from concentric annular regions. Uncertainties were derived from the
sky variance.  We find  differences between our values and those of Kirkpatrick et al. of typically 10\% for the fainter 3.6~$\mu$m data, and 5\% for the brighter  4.5~$\mu$m data. The differences are likely due to different data reduction techniques. Table 2 lists the {\it WISE} magnitudes as given by the WISE All-Sky Data Release\footnote {http://irsa.ipac.caltech.edu/Missions/wise.html} and the IRAC magnitudes as determined by us.
 
Figure 2 plots the difference between the magnitudes measured in the similar passbands IRAC [3.6] and {\it WISE} W1, and IRAC [4.5] and {\it WISE} W2; W1 extends the [3.6] bandpass to the blue, and W2 extends [4.5] to the red.  The difference between W2 and [4.5] magnitudes is small for T and Y dwarfs. W1 is significantly fainter than [3.6] for T dwarfs, because the bandpass extends into a region with very little flux. The single Y dwarf detected in W1 is WISE J1541$-$2250. The apparently 
low W1 $-$ [3.6] value for this object is likely to be spurious, as the  {\it WISE} images show no clear W1 source at the position of the W2 source, but does show other, bluer, nearby sources.  The T8.5 dwarf Wolf 940B (Burningham et al. 2009) also stands out in Figure 2. Leggett et al. (2010b) use near- and mid-infrared spectroscopy and photometry to show that Wolf 940B is a fairly typical late-type T dwarf with $T_{\rm eff} \approx 600$~K and $\log g \approx 5.0$  (cm s$^{-2}$). Most likely the {\it WISE} photometry for this object is compromised by the presence of the close and infrared-bright primary, as is suggested by the {\it WISE} images.

\section{Comparison to the Models}

\subsection{Description of the Models}

The models used here are described in detail in Morley et al. (2012). 
The model atmospheres are as described in Saumon \& Marley (2008) and 
Marley et al. (2002), with updates to the line list of NH$_3$ and of the 
collision-induced absorption of H$_2$ as described in Saumon et al. 
(2012). To these models, Morley et al. have added absorption and 
scattering by condensates of Cr, MnS, Na$_2$S, ZnS and KCl. These condensates  have been predicted to be present by Lodders (1999) and 
Visscher, Lodders \& Fegley (2006). Morley et al.  use the Ackerman \& 
Marley (2001) cloud model to account for these previously neglected 
clouds. The vertical cloud extent is determined by balancing upward 
turbulent mixing and downward sedimentation.  A parameter $f_{\rm sed}$ 
describes the efficiency of sedimentation, and is the ratio of the 
sedimentation velocity to the convective velocity; lower values of 
$f_{\rm sed}$ imply thicker  (i.e. more vertically extended) clouds.  We have found that our models
which include iron and silicate grains and  which have
$f_{\rm sed}$ of typically 2 -- 3 fit L dwarf spectra well, 
those with  $f_{\rm sed}$ 3 -- 4 fit T0 to T3 spectral types well, and 
cloud-free models fit T4 -- T8 types well (e.g. Stephens et al.
2009, Saumon \& Marley 2008). However for later spectral types significant 
discrepancies exist between our models and the observations in the near-infrared
(e.g. Leggett et al. 2009, 2012).

The new models with chloride and sulfide clouds show that these clouds are significant for dwarfs with 
$T_{\rm eff} = 400$ -- 900~K (approximately T7 to Y1 spectral types), with a peak impact at around 
600~K (or T9 types); Na$_2$S is the dominant species by mass, however at 400~K KCl is also important. 
A suite of models was generated with $400 \leq T_{\rm eff}$~K$ \leq 1300$, $4.0 \leq \log g \leq 5.5$ 
and $2 \leq f_{\rm sed} \leq 5$. Below $T_{\rm eff} \sim 400\,\rm K$ water clouds are expected to 
form, which have not yet been incorporated into the models (although water condensation is accounted 
for by removal of water from the gas opacity).

The chloride and sulfide clouds impact the 0.6 -- 1.3~$\mu$m wavelength region in particular, 
corresponding to the $zYJ$ photometric passbands. This otherwise clear region of the photosphere 
becomes opaque, and the flux emitted at these wavelengths then arises from a higher atmospheric layer 
that is cooler by around 200~K. The 1~$\mu$m flux is therefore significantly reduced by the presence 
of these clouds.

Morley et al. show that the new models which include the chloride and sulfide clouds reproduce the observations of late-type T dwarfs much better than is done by models without these clouds. In particular, for the well-studied 500~K brown dwarf UGPS J0722$-$0540, with data that covers optical to mid-infrared wavelengths, the SED is fit remarkably well with a  $T_{\rm eff} = 500$~K, $\log g = 4.5$, $f_{\rm sed} = 5$ model. The one region that is not fit well is the 5~$\mu$m region, where the models are too bright by  a factor of $\sim$ 2. This is because the cloudy models do not currently include departures from chemical equilibrium caused by vertical mixing.
The mixing enhances the abundance of CO and CO$_2$ and reduces the $5 \mu$m flux (Morley et al. 2012 and references therein).
It can be parametrized with an eddy diffusion coefficient of $K_{zz}$ cm$^2$ s$^{-1}$, where
values of log $K_{zz} = 2$ -- 6,  corresponding to mixing timescales of $\sim 10$ yr to $\sim 1$ hr, respectively, reproduce the observations of T dwarfs (e.g. Saumon et al. 2007). Leggett et al. (2012) find that UGPS J0722$-$0540 is undergoing vigorous mixing, with log $K_{zz} \approx 5.5$ -- 6.0,
 which results in an increase of W2 by $\gtrsim 0.3$ magnitude (their Figures 6 and 7).
Cloudless equilibrium and non-equilibrium models imply that the impact on the W2 or [4.5] magnitudes is around
$0.3 \pm 0.1$ magnitudes for  $600 \leq T_{\rm eff}$~K $\leq 800$ and $4.0 \leq \log g \leq 5.0$. Given that the impact for the 500~K dwarf UGPS J0722$-$0540 is of a similar order, and that CO is dredged up into atmospheres as cool as Jupiter's (e.g. Noll et al. 1988), for simplicity here we add 0.3 magnitudes to the  W2 and [4.5] magnitudes computed for all the model sequences used in this paper, at all values of $T_{\rm eff}$, to mimic the effect of vertical mixing.
We find below that such a correction reproduces the observed trends in T and Y dwarf W2 colors quite well (see \S 4.3). Our next generation  of models will incorporate water clouds and mixing.

Tables 3 and 4 list MKO-system near-infrared photometry, and IRAC and {\it WISE} photometry, generated by the Morley et al. (2012) cloudy models and the Saumon et al. (2012) cloud-free models, for brown dwarfs at 10~pc. The cloudy models include Na$_2$S, MnS, ZnS, Cr, KCl condensate clouds and do not include Fe, Mg$_2$SiO$_4$, or Al$_2$O$_3$ condensate clouds.  Colors are calculated using Saumon \& Marley (2008) cloud-free evolution model grids, and all are in the Vega system.

\subsection{Data Sources}

Our sample for this work consists of brown dwarfs with spectral type T6 and later, that have been detected in the W2 band, and that have MKO-system near-infrared photometry. Eighty-three brown dwarfs satisfy these criteria at the time of writing. The Appendix contains a data table which gives, for these 83 brown dwarfs: coordinates, spectral type, distance modulus, $izYJHKL^{\prime}M$[3.6][4.5][5.8][8.0]W1W2W3W4 photometry, uncertainties in distance modulus and photometry, and source references.
 
The data table also flags objects that are close binary systems. Recently, four new binary systems have been identified at or near the T/Y dwarf boundary (Burgasser et al. 2012; Liu et al. 2011, 2012). These systems are extremely useful for testing the models, as they will have the same metallicity, and the same age. The age constraint, together with the measured luminosity, translates into a maximum difference in $\log g$ of 0.4 dex. The binary systems are listed in Table 5. Burgasser et al. and Liu et al. give resolved MKO-system near-infrared photometry for the binaries. The systems are not resolved by IRAC or WISE. We have estimated W2 magnitudes for the individual components using the unresolved values, and the expected difference between the components based on the difference in spectral types (Figure  3). The uncertainty in the estimate is around 0.3 magnitudes which is derived from the range of W2 values that will produce the measured unresolved W2.

\subsection{Color Sequences}

Figure 3 is a plot of various colors as a function of spectral type. Spectral types are from Kirkpatrick et al. (2011, 2012), and include the adjustment to the late-type T dwarf classification described in Cushing et al. (2011).  The reader should note that the definition of the end of the T sequence and the start of the Y sequence is very preliminary at this time. It is likely that the classification scheme will need to be revised once more Y dwarfs are known. $Y - J$ for the Y dwarfs shows a marked decline, while  $J - H$ is mostly flat up to the Y2 class. $H - K$ is also mostly flat, although there is more scatter, possibly reflecting a range in gravity, or age, for the sample (see discussion of Figure 8 below). $H -$ W2 shows a steady and rapid increase with later spectral types. [3.6] $-$ W2 and W2 $-$ W3 also show an increase, however the object classified as Y2 appears to turn over in the [3.6] $-$ W2 diagram, and the W2 $-$ W3 diagram shows a lot of scatter (although the uncertainties are quite large). We come back to the Y2 dwarf  WISEPC J1828$+$2650 in \S 4.5.

Figure 4 shows absolute $J$ magnitude as a function of the near-infrared colors $Y - J$, $J - H$ and $J - K$. 
Photometry and parallaxes are taken from the sources listed in the Appendix.
Model sequences are shown, both cloud-free and cloudy, for $f_{\rm sed} = 3$ and 5, and for a range in gravity between $\log g = 4.0$ and 5.0. At the temperatures of primary interest here,  $300 \leq T_{\rm eff}$~K$ \leq 600$, $\log g = 4.0$ corresponds to a mass around 5 M$_{\rm Jupiter}$ and age 0.1 -- 1 Gyr, $\log g = 4.5$ to mass  12  M$_{\rm Jupiter}$ and age 1 -- 10 Gyr, and $\log g = 5.0$ to mass  30  M$_{\rm Jupiter}$ and age $>$ 6 Gyr. Given that the latest-type T dwarfs and the Y dwarfs are necessarily nearby, we expect them to have an age similar to the Sun, and therefore the  $\log g = 4.5$ model sequence should be most representative of the sample (solid curves in Figure 4). It can be seen that the new cloudy models are required in order to reproduce the observed significant reddening of around 1 magnitude in $J - H$  for brown dwarfs with  $T_{\rm eff} = 400$ -- 800~K. In the $J - K$ diagram the datapoints occupy a wider color envelope, possibly reflecting an intrinsic scatter in metallicity or gravity, which would impact the strong H$_2$ opacity in the $K$ band. The cloudy models do not reproduce the $Y - J$ colors at $T_{\rm eff} \approx 400$~K, possibly due to changes associated with the formation of water clouds. 

Figure 5 shows absolute W2 magnitude as a function of the mid-infrared colors $J -$ W2, $H -$ W2 and W2 $-$ W3.   W3 is a wide filter spanning 7.5 -- 16.5 $\mu$m. Only around 5 known late-type T dwarfs (and no Y dwarfs) are detected in the W4 filter (19.8 -- 25.5 $\mu$m). In these plots, as explained above, we have added 0.3 magnitudes to the W2 magnitudes calculated by both the cloudy and cloud-free  models, to mimic the affect of vertical mixing. With this adjustment, the models fit the data quite well, although the  $H -$ W2 diagram suggests that the correction should be larger for both
cloudy and cloud-free models, and the  $J -$ W2 diagram suggests that the correction should be larger for the cloudy models.
The two unusually red objects at M$_{\rm W2} \approx 13$ are both metal-poor high-gravity late-type T dwarfs,  2MASS J09393548$-$2448279 (Burgasser et al. 2008, Leggett et al. 2009) and SDSS 1416$+$1348B (Scholz 2010a, Burningham et al. 2010b); the former is  likely to be a binary system. Metal-poor T dwarfs are bright at 4.5 $\mu$m  (e.g. Leggett et al. 2010a); non-solar metallicities have not yet been incorporated into these models.
WISE J1738$+$2732 appears to be very red in W2 $-$ W3. This object is also unusually blue in $Y - J$ (Figures 4 and 7) and warrants further study.  WISEPC J1828$+$2650 also stands out in Figure 5 --- we return to this in \S 4.5. 

$Y - J$ as a function of $J - H$ and $J -$ W2 is shown in Figures 6 and 7 respectively. Model sequences are also shown, as described above. The impact of the chloride and sulfide clouds is  apparent in Figure 6, where cloud-free models extend to $J - H$ colors that are much bluer than observed. In Figure 6 we have included data from the UKIDSS database to illustrate the location of the general stellar population. It will be extremely difficult to identify Y dwarfs in sky surveys using near-infrared colors alone. These cold brown dwarfs have now wrapped around in $YJH$ colors to occupy a very similar region to that occupied by warm  stars. Allowing for uncertainties in the survey data at faint limits, it is not possible to extract even the earliest Y dwarfs, with $Y - J \sim 0$ and  $J - H \sim  -0.4$. Figure 7 shows, similarly to Figure 4, that the cloudy models do not reproduce the $Y - J$ colors at $T_{\rm eff} \approx 400$~K.

Figure 8 shows $H -$ W2 as a function of $J - H$, $H - K$ and [3.6] $-$ W2. Model sequences are also shown, as described above. Again, the new cloudy models nicely account for the reddening of $J - H$.  There is  a degeneracy 
between the effect of $f_{\rm sed}$ and $\log g$ in the $J - H$ diagram,
where an increase in  $f_{\rm sed}$  can be compensated by a decrease in gravity. The $H - K$ diagram, however, separates out the gravity and  $f_{\rm sed}$ sequences, allowing us to differentiate between the two parameters. Although $K$ is very faint for these objects it will be worth expending some telescope time to obtain such data in order to more fully understand this new class of objects.

The model comparisons that we show in Figures 4, 5, 7 and 8 allow us to estimate the properties of the five Y0 -- Y0.5 dwarfs for which we present new photometry in this paper. We discuss the Y2 dwarf in \S 4.5. We find that WISEP J0410+1502,  WISEP J1738$+$2732  and WISEPC J2056$+$1459 have $T_{\rm eff} \approx 400$ -- 450~K, $\log g \approx 4.5$ and  $f_{\rm sed} \approx 3$ (based on the $J - H$ and $H - K$ colors).  WISEPC J1405$+$5534 and  WISEPC J1541$-$2250 are cooler than covered by the cloudy models, but the cloud-free models indicate  
$T_{\rm eff} \approx 350$~K and 300 -- 350~K respectively. Trends in the figures indicate that WISEPC J1405$+$5534 has a higher gravity than WISEPC J1541$-$2250, and we estimate that 
$\log g \approx 4.5$ and 4.0 -- 4.5 for WISEPC J1405$+$5534 and  WISEPC J1541$-$2250, respectively. These temperatures and gravities, with implied ranges in mass and age, are listed in Table 6. We have assumed solar metallicity, which for this sample within 10 pc of the Sun is plausible. 

The comparisons can also be tested against the binary parameters given in Table 5. The model sequences are consistent with a single-age solution for the three pairs:   WISE J1217$+$1626AB, CFBDS 1458$+$1013AB and WISE J1711$+$3500AB. WISE J0458$+$6434AB appears to be a very similar pair of T9s which we cannot constrain. For all three binaries the models support the older 5~Gyr solution and do not support the younger 1~Gyr solution, based on  agreement with the higher gravity sequences in Figure 8. The 5~Gyr solution for WISE J1217$+$1626  implies $\log g = 5.0$ and 4.7 for the primary and secondary, respectively (Table 5). The middle panel of Figure 8 shows that the observations are consistent with the  $f_{\rm sed} = 5$  (i.e. relatively thin cloud layers) $\log g = 5.0$ and 4.5 sequences, and that $H - K$ is too blue to be consistent with with any lower-gravity solution.  Similarly CFBDS 1458$+$1013A constrains the system to the older solution where  $\log g = 5.0$ and 4.6 for the primary and secondary, respectively. The colors of the secondary are consistent with this, although they do not constrain the solution. For  CFBDS 1458$+$1013A the clouds appear to be very thin to non-existent. The 5~Gyr solution for  WISE J1711$+$3500AB  has  $\log g = 5.2$ and 4.8 for the primary and secondary, respectively. WISE J1711$+$3500B constrains the system to this older solution, as $H - K$ is too blue to be consistent with the alternative 1~Gyr  $\log g = 4.3$ solution, for any cloud parameter.  WISE J1711$+$3500B  also has a thin cloud layer with  $f_{\rm sed} \gtrsim 5$.  WISE J1711$+$3500A is too warm to constrain the gravity  of the system.

\subsection{Spectral Energy Distributions}

The color sequences indicate that WISEPC J2056$+$1459 has $T_{\rm eff} = 400$ -- 450~K, $\log g = 4.5$ and  $f_{\rm sed} = 3$. Cushing et al. (2011) present $J$-band and $H$-band spectra for this Y0 dwarf, which we have flux calibrated using our photometry. Figure 9 combines this spectrum with the far-red spectrum obtained here, and compares these spectra to models with  $T_{\rm eff} = 400$~K and  $\log g = 4.5$, with and without clouds. It can be seen that the cloudy model provides a superior fit in the red and at $K$. The discrepancy at 1.0~$\mu$m and 1.5~$\mu$m, where the cloudy model flux is fainter than observed, can be explained by an overly-strong NH$_3$ absorption in the models, due to the neglect of vertical mixing (see \S 4.1). The mixing enhances N$_2$ while decreasing the abundance of NH$_3$. This effect can also explain the lack of a strong NH$_3$ doublet at 1.03 $\mu$m in Figure 1.   The discrepancy at 1.60 -- 1.65 ~$\mu$m, where the model flux is too high, is most likely due to remaining incompleteness in the CH$_4$ opacity line  list (CH$_4$ is the dominant opacity at 1.6~$\mu$m).

The color sequences suggest that WISEPC J1541$-$2250, the coldest dwarf in the sample (apart from WISEPC J1828$+$2650, see below),
 has colors approaching those of the cloud-free models (Figures 4, 7, 8).
In Figure 10 we plot the near-infrared spectrum of this brown dwarf and compare it to a cloud-free model with $T_{\rm eff} = 325$~K and $\log g = 4.5$.  The cloudless synthetic spectrum reproduces the 1.0 -- 1.6 $\mu$m data quite well, with the caveats as before of overly strong NH$_3$  due to the neglect of mixing. 

The fact that a cloudless model fits the spectrum reasonably well despite the fact that water clouds are expected by $325\,\rm K$ may not be surprising.  For such a gravity near $T_{\rm eff}\sim 300\,\rm K$, the cloud base forms well above the 1 bar level where there is less mass available to condense, compared to still lower effective temperatures where the cloud base is much deeper (e.g. in Jupiter).  Furthermore, ice crystals or water drops with radii less than about $0.5\,\rm \mu m$ do not interact strongly with near-infrared photons since the relevant Mie absorption and scattering efficiencies are very low (see Figure 3 of Zsom et al. 2012 for example).  When water clouds are found deeper in the atmosphere or form with larger radii we would expect to see a greater effect.  In the near-infrared this would likely first become noticeable in the $Y$- and $J$-bands because of the longer atmospheric pathlength in these opacity windows, and the greater influence of scattering at the shorter wavelengths. Burrows, Sudarsky \& Lunine (2003) do include water condensates in their models of brown dwarfs with 
$130 \leq T_{\rm eff}$~K$\leq 800$. They derive condensate particle sizes of 20 to 150 $\mu$m, and nevertheless still find that the absorptive opacity of the water clouds is small, such that water ice clouds only have a  secondary influence on the spectra of the coolest isolated brown dwarfs. Burrows et al. also calculate that NH$_3$ clouds form when  $T_{\rm eff} \lesssim 160$~K, i.e. cooler than the {\it WISE} Y dwarfs. Further modelling is required to investigate cloud formation for brown dwarfs with  $T_{\rm eff} \lesssim 400$~K.

\subsection{WISEPC J1828$+$2650}

WISEPC J1828$+$2650 has the reddest $H -$ W2 color of any known Y dwarf
($H -$ W2 $= 8.46 \pm  0.25$), 
and is therefore likely to be the coolest. Cushing et al. classify this 
object as Y2. Figures 4 and 5 show that this Y2 dwarf is as 
intrinsically bright in the near-infrared as the Y0.5 WISEPC 
J1541$-$2250, and is actually brighter in the mid-infrared. 
WISEPC J1828$+$2650 is also similar in luminosity to the Y0  WISEPC J1405$+$5534.
In order  
for a cooler brown dwarf to be more luminous than a warmer 
one, the radius must be significantly larger ($T_{\rm eff}^4$ 
is compensated by the factor $R^2$). In this particular case, 
with an estimated 10\% difference in temperature, the radii would need 
to differ by $\gtrsim$ 20\%. Assuming the rest of the 350 -- 450~K sample
has a typical age of a few Gyr and $\log g \approx 4.5$, then the radius
of WISEPC J1828$+$2650 would need to be $\gtrsim 0.126~ $R$_{\odot}$.
This in turn implies an age younger than 50~Myr, and mass smaller than
1 M$_{\rm Jupiter}$.

While the young age and low mass are possible, we have explored an
alternative solution of binarity for WISEPC J1828$+$2650.
In this scenario, the cooler component would be almost as bright as the 
warmer component in the mid-infrared, but be significantly fainter in 
the near-infrared. We determined solutions that 
(i) reproduced the observed flux when the component fluxes were combined, 
(ii) produced absolute magnitudes generally consistent with our models
(as the distance is known, Table 2), and 
(iii) had $T_{\rm eff}$ and $\log g$ parameter pairs consistent 
with a coeval pair.  Table 7 gives our 
synthetically resolved photometry for the components;
the uncertainty is  0.3 magnitudes, constrained by the requirement to 
reproduce the observed flux. Figure 3 shows the proposed components on
spectral type:color diagrams --- trends with type are sensible if this brown dwarf is a binary.
Figures 4 to 8 show the proposed components on
color diagrams. Extrapolating from the models and the observed sequences,
the proposed binary appears to consist of  
a $T_{\rm eff} \approx 325$~K and $\log g \lesssim 4.5$ 
primary, and a $T_{\rm eff} \approx 300$~K and $\log g \gtrsim 4.0$ 
secondary. These imply a system age of around 2~Gyr, and component  masses of 
around 10 and 7 M$_{\rm Jupiter}$. 300~K brown dwarfs appear to be
significantly brighter than the cloud-free models
in the $K$ and W2 bands (Figures 4, 5 and 10), a discrepancy to be addressed by the next generation of models.

\subsection{Conclusion}

We have obtained $YJHK$, or a subset, for the six objects discovered in the {\it WISE} database that have been identified
as Y dwarfs by Cushing et al. (2011). We find large differences of 0.5 -- 1.0 magnitudes  between our MKO-system magnitudes and those published
by Cushing et al. and Kirkpatrick et al. (2012) for three of 
the brown dwarfs at $J$ or $H$. As our photometry is consistent with the  SEDs
published by  Cushing et al. and Kirkpatrick et al. we suspect that the previously published Palomar WIRC photometry is in error for
these three objects at those particular wavebands. Our results are otherwise consistent with the previously published data, where they overlap.

We have also obtained a far-red spectrum of the Y0 dwarf  WISEPC J2056$+$1459, which has $T_{\rm eff} \approx 400$~K. The spectrum shows that 
the Cs I lines seen in the 500~K brown dwarf  UGPS J0722$-$0540 (Leggett et al. 2012) are not seen in this cooler object. This is not unexpected, as at these temperatures Cs should exist predominantly in the form of CsCl (Lodders 1999).

We confirm here that new models which include
clouds of  Cr, MnS, Na$_2$S, ZnS and KCl condensates reproduce the near-infrared colors of the latest-type T dwarfs, and the earliest-type Y dwarfs, quite well, as previously shown by Morley et al. (2012).  The $J - H$:$H -$ W2 and $H - K$:$H -$ W2 color-color diagrams can be used to estimate  $T_{\rm eff}$, $\log g$ and sedimentation efficiency    $f_{\rm sed}$ for the earliest Y dwarfs (Figure 8).
We find that WISEP J0410+1502,  WISEP J1738$+$2732  and WISEPC J2056$+$1459 have $T_{\rm eff} \approx 400$ -- 450~K, $\log g \approx 4.5$ and  $f_{\rm sed} \approx 3$.  We also find that WISEPC J1405$+$5534 and  WISEPC J1541$-$2250 are cooler than currently covered by the cloudy models, with $T_{\rm eff} \approx 350$~K and 300 -- 350~K and $\log g \approx 4.5$ and 4.0 -- 4.5, respectively. We find that the 1.0 -- 1.6 $\mu$m spectrum of  WISEPC J1541$-$2250 can be reproduced quite well with a cloud-free $T_{\rm eff} = 325$~K $\log g = 4.5$ model, which is somewhat surprising, as water clouds  would be expected to form in such cool atmospheres. The lack of a strong cloud signature may be due to the fact that the cloud base for these relatively low-gravity dwarfs lies high in the atmosphere, and small droplets have low absorption and scattering efficiencies.  Improved water cloud models are clearly required.

The temperatures and gravities of the five Y0 -- Y0.5 dwarfs imply a mass range of 5 -- 15 M$_{\rm Jupiter}$ and an age around 5~Gyr. We determine similar ages for three binary systems composed of late-T and early-Y dwarfs (Liu et al. 2011, 2012). In each case comparison to model sequences indicates a value of $\log g$ which corresponds to an age around 5 Gyr. The T9 -- Y0 components of the three binaries --- WISE J1217$+$1626A and B, CFBDS 1458$+$1013A and B, and WISE J1711$+$3500B -- appear to have thinner clouds than the Y0 dwarfs  WISEP J0410+1502,  WISEP J1738$+$2732  and WISEPC J2056$+$1459:  $f_{\rm sed} \gtrsim 5$ compared to  $f_{\rm sed} \approx 3$.

We have found that vertical mixing is likely to be important in the atmospheres of the early-type 400~K Y dwarfs,
as it is in the atmospheres of T dwarfs, and indeed as it is for Jupiter (e.g. Noll, Geballe \& Marley 1997, Leggett et al. 2002, Geballe et al. 2009). The mixing enhances the abundance of N$_2$ at the expense of NH$_3$, explaining the lack of detection of what would otherwise be strong NH$_3$ absorption at 1.03 $\mu$m and 1.52 $\mu$m  (Figures 1, 9, 10). 

The object WISEPC J1828$+$2650 is peculiar. Assuming that the parallax for  WISEPC J1828$+$2650 is not in error, its luminosity is not consistent with its colors, unless it is younger than around 50 Myr.  A more plausible solution may be that  WISEPC J1828$+$2650 is a binary system. We find that a $T_{\rm eff} \approx 325$~K and $\log g \lesssim 4.5$ primary, and a $T_{\rm eff} \approx 300$~K and $\log g \gtrsim 4.0$ 
secondary,  would follow the observational trends seen with type, absolute magnitude and color (Figures 3 to 8). The brightness and colors of the proposed components are also consistent with the cloud-free models, with the exception that the model fluxes at $K$ and W2  are too faint by 0.5 -- 1.0 magnitudes. 

In future work we will incorporate water clouds into the models, as well as mixing. The atmospheres of these 300 -- 500 K objects are extremely complex, but nevertheless the models, which now include chloride and sulfide clouds, have allowed us to estimate temperature and gravity, and hence mass and age, for these exciting {\it WISE} discoveries. We look forward to the discovery of more  5 -- 10 M$_{\rm Jupiter}$ objects in the solar neighborhood.

\acknowledgments

DS is supported by NASA Astrophysics Theory grant NNH11AQ54I.
Based on observations obtained at the Gemini Observatory, which is operated by the Association of Universities for
 Research in Astronomy, Inc., under a cooperative agreement with the
    NSF on behalf of the Gemini partnership: the National Science
    Foundation (United States), the Science and Technology Facilities
    Council (United Kingdom), the National Research Council (Canada),
    CONICYT (Chile), the Australian Research Council (Australia),
    Minist\'{e}rio da Ci\^{e}ncia, Tecnologia e Inova\c{c}\~{a}o (Brazil)
    and Ministerio de Ciencia, Tecnolog\'{i}a e Innovaci\'{o}n Productiva
    (Argentina). SKL's research is supported by Gemini Observatory.  This publication makes use of data products from the Wide-field Infrared Survey Explorer, which is a joint project of the University of California, Los Angeles, and the Jet Propulsion Laboratory/California Institute of Technology, funded by the National Aeronautics and Space Administration. This research has made use of the NASA/ IPAC Infrared Science Archive, which is operated by the Jet Propulsion Laboratory, California Institute of Technology, under contract with the National Aeronautics and Space Administration.

\appendix

\section{Data Table}

Table 7 presents a sample of the data used in this paper. The full dataset is available online. The table gives, for 83 brown dwarfs
which have spectral types of T6 or later,  that have been detected in the W2 band, and that have MKO-system near-infrared photometry: coordinates, spectral type, distance modulus, $izYJHKL^{\prime}M$[3.6][4.5][5.8][8.0]W1W2W3W4 photometry, uncertainties in distance modulus and photometry, and source references. Close binary systems are also indicated. The coordinates are for equinox 2000, please note that epoch varies and the dwarfs can have high proper motion. The $iz$ photometry is on the AB system, while the rest of the photometry is on the Vega magnitude system. The $YJHKL^{\prime}M$ are on the MKO photometric system, with $Y$ data obtained using NIRI reduced by 0.17 magnitudes to put it on the UKIDSS $Y$-system. The uncertainty in the spectral types are not given, but is typically 0.5 of a subclass, except for the spectral types latern than T8. It is likely that the classification scheme for those objects will have to be revised as more Y dwarfs are found, and we estimate an uncertainty of 1 subclass in type. All {\it WISE} photometry is taken from the {\it WISE} All-Sky Data Release \footnote {http://irsa.ipac.caltech.edu/Missions/wise.html}.

The online table contains 48 columns:\\
1. Name\\
2. Other names(s)\\
3. R.A. as HHMMSS.SS\\
4. Declination as SDDMMSS.S\\
5. Spectral type\\
6. Distance modulus as $M - m$ magnitudes\\
7. Absolute $J$ magnitude\\
8. $i$\\
9. $z$\\
10. $Y$\\
11. $J$\\
12. $H$\\
13. $K$\\
14. $L^{\prime}$\\
15. $M$\\
16. IRAC [3.6]\\
17. IRAC [4.5]\\
18. IRAC [5.8]\\
19. IRAC [8.0]\\
20. W1\\
21. W2\\
23. W3\\
24. W4\\
25. Uncertainty in  $M - m$\\
26. Uncertainty in $i$\\
27. Uncertainty in $z$\\
28. Uncertainty in $Y$\\
29. Uncertainty in $J$\\
30. Uncertainty in $H$\\
31. Uncertainty in $K$\\
31. Uncertainty in $L^{\prime}$\\
31. Uncertainty in $M$\\
32. Uncertainty in [3.6]\\
33. Uncertainty in [4.5]\\
34. Uncertainty in [5.8]\\
35. Uncertainty in [8.0]\\
36. Uncertainty in W1\\
37. Uncertainty in W2\\
38. Uncertainty in W3\\
39. Uncertainty in W4\\
40. Reference for binarity\\
41. Reference for discovery\\
42. Reference for spectral type\\
43. Reference for trigonometric parallax\\
44. Reference for $iz$\\
45. Reference for $Y$\\
46. Reference for $JHK$\\
47. Reference for $L^{\prime}M$\\
48. Reference for IRAC photometry\\

Data are taken from this work, Burningham et al. 2013 (in preparation), data release 9 of both the UKIDSS and SDSS catalogs, and the {\it WISE} All-Sky Data Release. Brown dwarf discoveries and other data are taken from the following publications: Burgasser et al. 1999, 2000, 2002, 2003a and b, 2004, 2006a and b, 2008, 2012; Burningham et al. 2008, 2009, 2010a and b, 2011; Chiu et al. 2006; Cushing et al. 2011; Dahn et al. 2002; Delorme et al. 2008a and b, 2010; Dupuy \& Liu 2012; Faherty et al. 2012; Geballe et al. 2001; Golimowski et al. 2004; Harrington \& Dahn 1980; Hewett et al. 2006; Kirkpatrick et al. 2011, 2012; Knapp et al. 2004; Leggett et al. 2002, 2007, 2009, 2010, 2012; Liu et al. 2011, 2012; Lodieu et al. 2007, 2009; Lucas et al. 2010; Marocco et al. 2010; Patten et al. 2006; Pinfield et al. 2008, 2012; Scholz 2010a and b; Strauss et al. 1999; Tinney et al. 2003, 2005; Tsvetanov et al. 2000; Vrba et al. 2004; Warren et al. 2007.



\clearpage

\begin{figure}
\includegraphics[angle=-90,scale=.6]{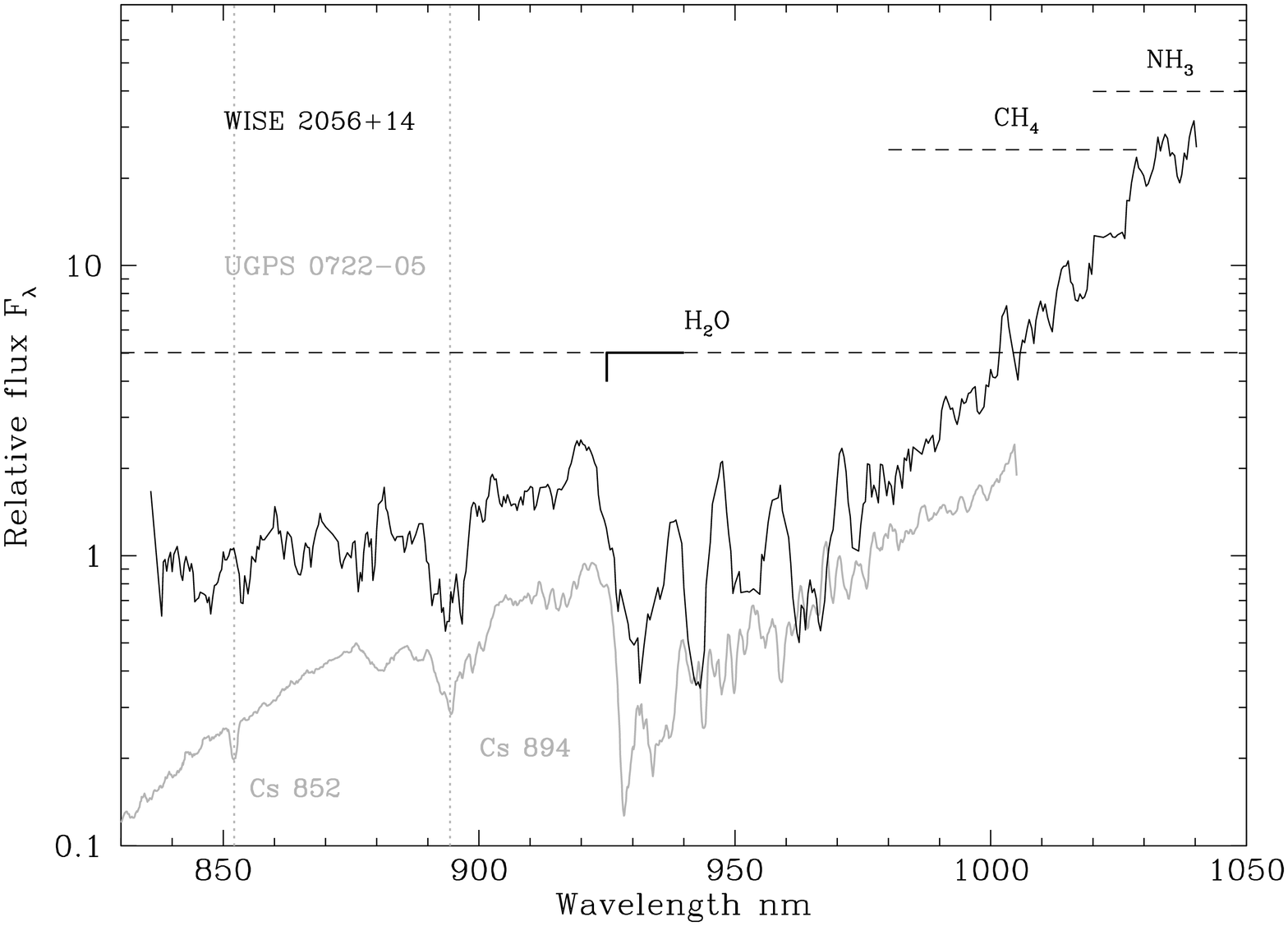}
\caption{Observed GMOS spectra of WISE J2056$+$1459 (black curve) and 
UGPS 0722$-$05 (gray curve), from this work and Leggett et al. (2012) 
respectively. The spectrum of WISE J2056$+$1459 has been smoothed with a 
5 pixel boxcar. Principal absorbers are indicated. H$_2$O absorption occurs throughout 
the wavelength region shown, however the strong bandhead at 0.93 $\mu$m is indicated.
Cs features seen  in UGPS 0722$-$05 are not seen in  WISE J2056$+$1459.
\label{fig1}}
\end{figure}

\begin{figure}
\includegraphics[angle=-90,scale=.6]{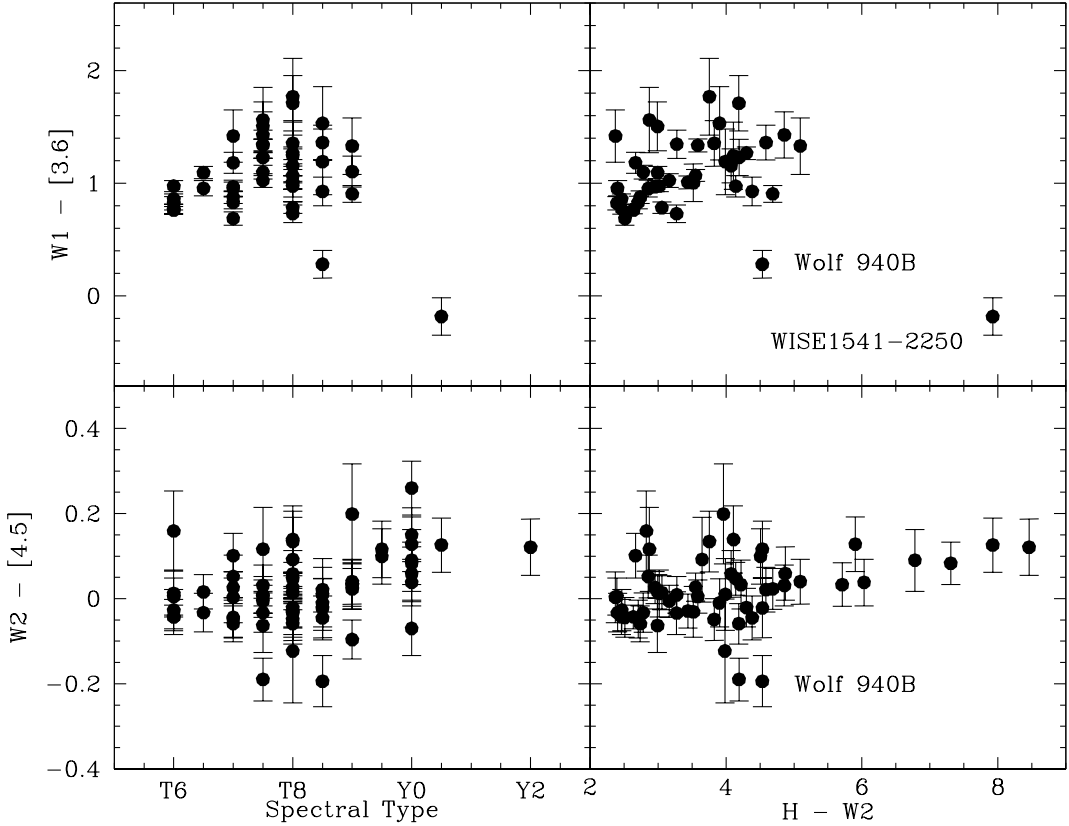}
\caption{Measured differences 
between photometry obtained with the 
similar passbands IRAC [3.6] and {\it WISE} W1, and IRAC [4.5] and {\it WISE} W2, as 
a function of spectral type and $H -$ W2 color. Spectral types for the 
T9 and Y dwarfs are from Kirkpatrick et al. (2011, 2012), and include 
the adjustment to the late-type T dwarf classification described in 
Cushing et al. (2011). Data sources are described in the Appendix.  The photometry for Wolf 940B is likely compromised by the 
presence of the nearby very infrared-bright primary. The low W1 $-$ [3.6] value for WISE J1541$-$2250 is likely to be spurious,
as the  {\it WISE} images show no clear W1 source at the position of the W2 source, but does show other, bluer, nearby sources. 
\label{fig2}}
\end{figure}

\begin{figure}
\includegraphics[angle=-90,scale=.6]{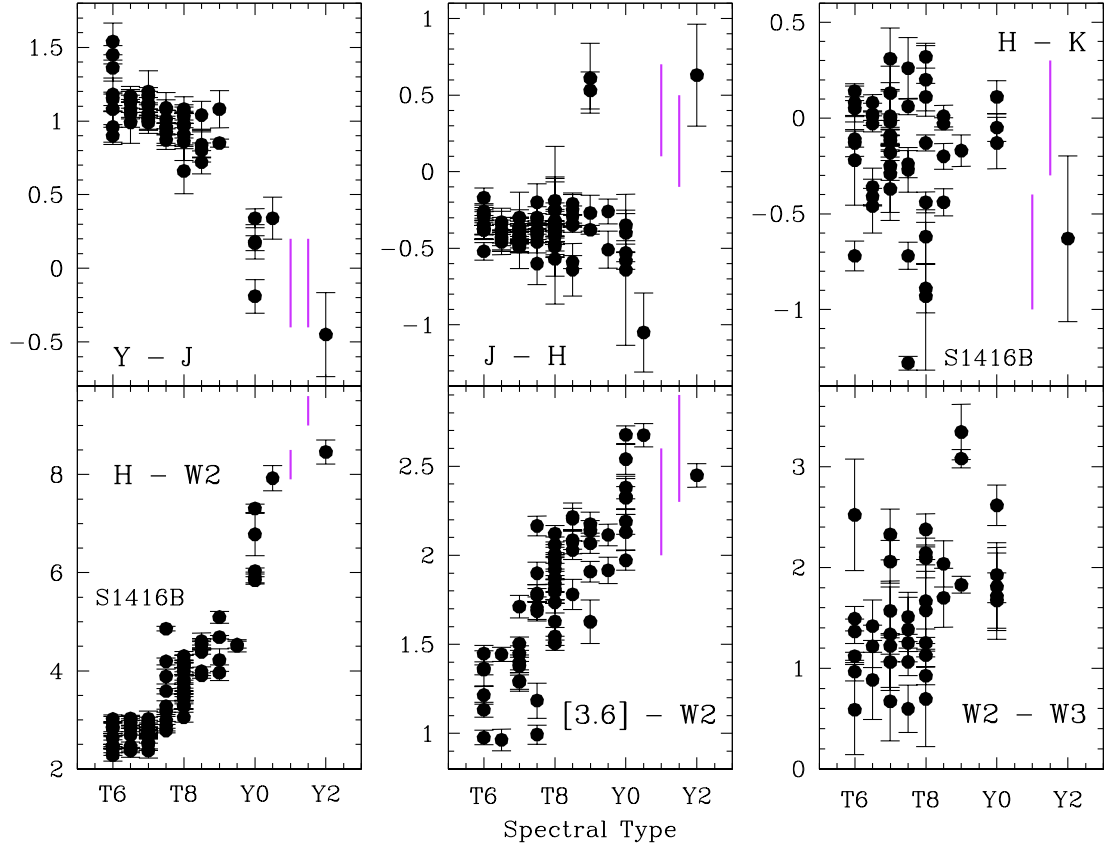}
\caption{Colors as a function of spectral type. Spectral types are as in 
Figure 2. Data sources are described in the Appendix. The 
IRAC [3.6] magnitude is plotted instead of  {\it WISE} W1 as the 
uncertainties are generally smaller. Figure 2 shows the difference 
between [3.6] and W1 as a function of type and color. The object that is very 
blue in $H - K$ and red in $H -$ W2 is the very metal-poor high-gravity 
T dwarf SDSS 1416$+$1348B (Scholz 2010a, Burningham et al. 2010b). Two T9 dwarfs stand out as red in $J - H$: 
WISEJ 1614$+$1739 and WISEJ 2325$-$4105. The photometry is taken from Kirkpatrick et al. (2011),
and should be repeated to exclude the possiblity of measurement error, or error in photometric systems. 
Vertical lines (violet in the online version) indicate the colors of 
WISEPJ 1828$+$2650 if it consists of a 325~K and 300~K binary (see \S 4.5); for clarity we assign types of Y1 and Y1.5.
\label{fig3}}
\end{figure}

\begin{figure} \includegraphics[angle=0,scale=.6]{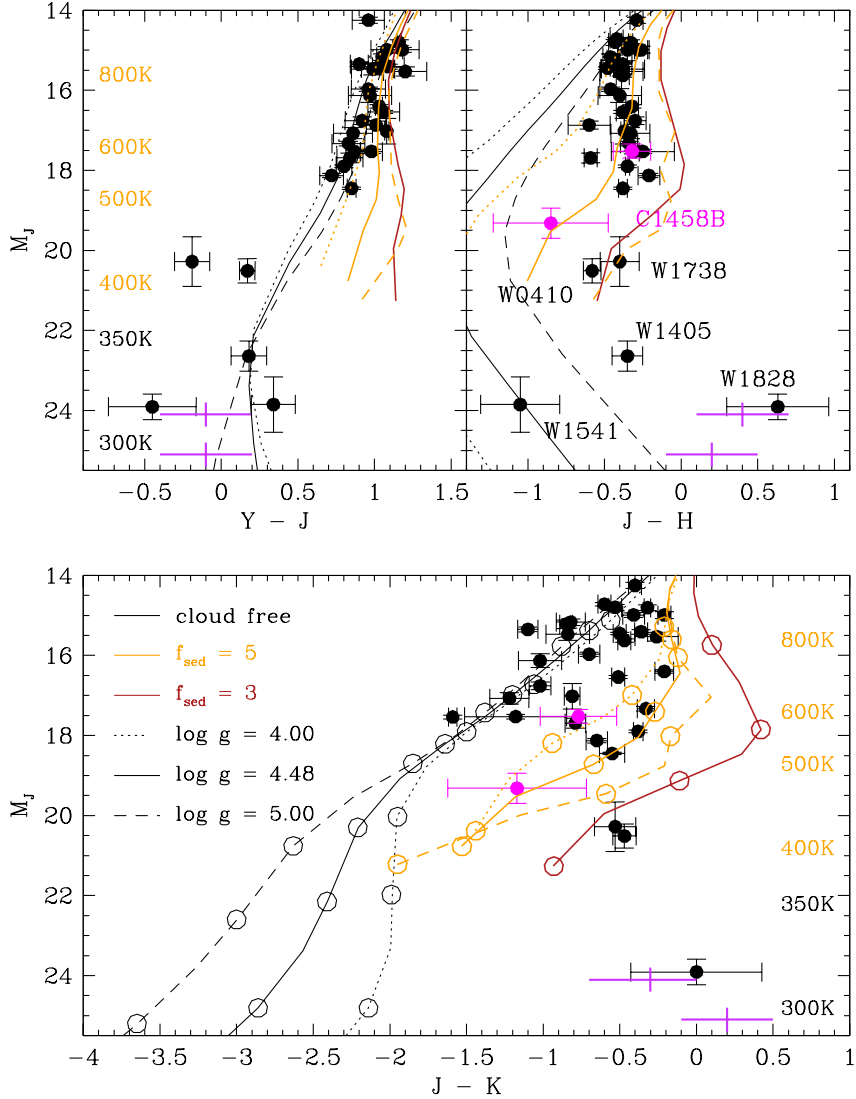} 
\caption{Absolute $J$ magnitude as a function of $Y - J$, $J - H$ and 
$J - K$. Crosses without datapoints (violet in the online version) 
indicate the colors of WISEPJ 1828$+$2650 if it consists of a 325~K and 
300~K binary (see \S 4.5). The pair of lighter datapoints (magenta in 
the online version) represent CFBDS 1458$+$1013A and B. Curves are 
model sequences, as described in the legend. In the lower panel large 
open circles along the sequences indicate where $T_{\rm eff} =$ 300, 
350, 400, 500, 600 and 800~K. $T_{\rm eff}$ values on the vertical axes 
correspond to the $f_{\rm sed} = 5$ and $\log g = 4.5$ model for $400 
\leq T_{\rm eff}$~K$ \leq 800$ and the cloud-free models for $300 \leq 
T_{\rm eff}$~K$ \leq 350$. All data are on the MKO photometric system. 
\label{fig4}} \end{figure}

\begin{figure}
\includegraphics[angle=0,scale=.65]{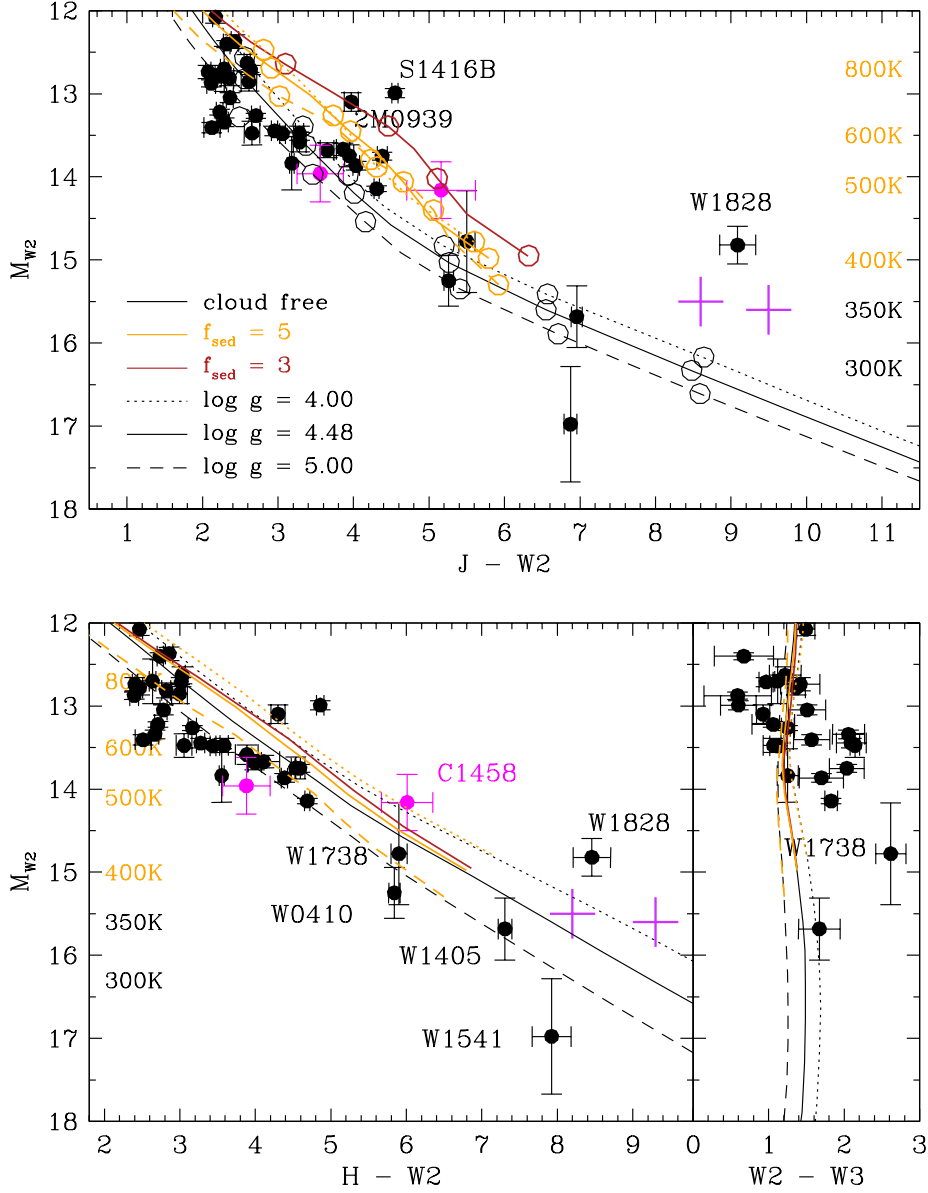}
\caption{Absolute W2 magnitude as a function of the mid-infrared colors 
$J -$ W2, $H -$ W2 and W2 - W3. Symbols and curves are as in Figure 4. 
The model W2 values have been increased by 0.3 magnitudes to mimic the effect
of  non-equilibrium chemistry  as  described in the text.
\label{fig5}}
\end{figure}

\begin{figure}
\includegraphics[angle=0,scale=.6]{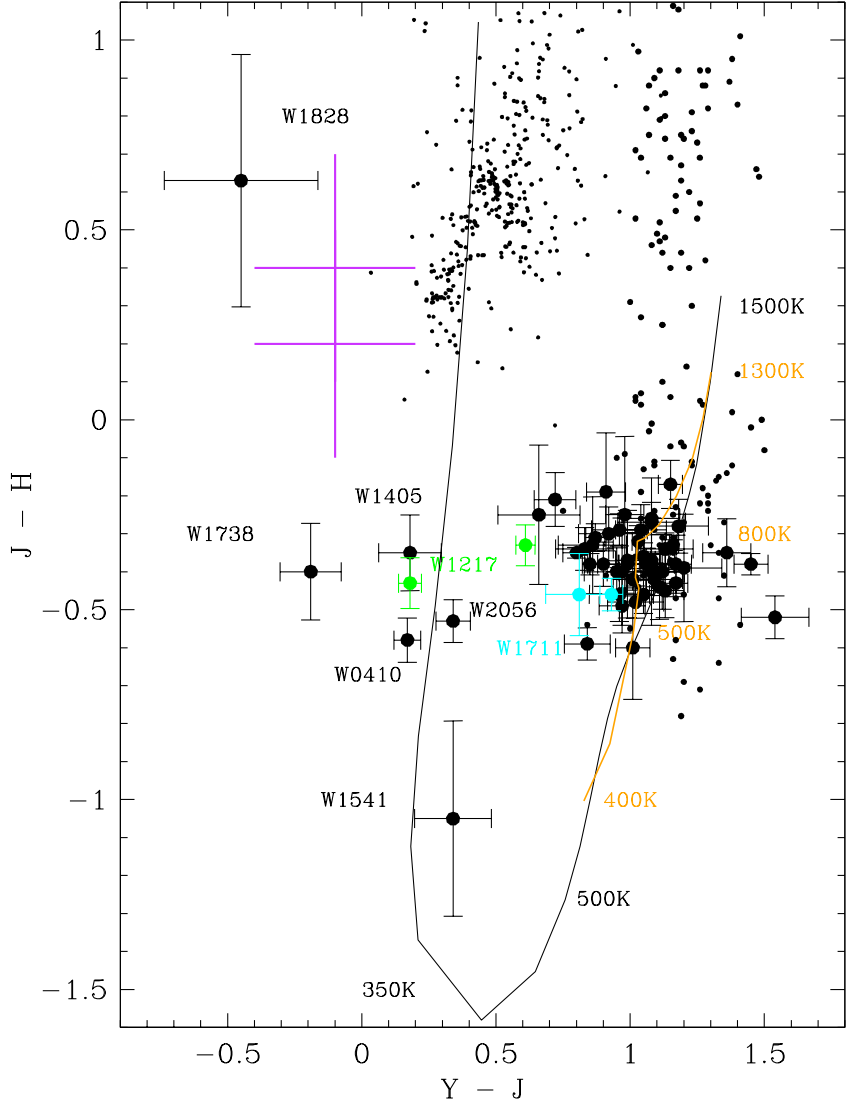}
\caption{$J - H$ as a function of $Y - J$. Crosses (violet in the online version) indicate the colors of 
WISEPJ 1828$+$2650 if it consists of a 325~K and 300~K binary (see \S 4.5). Curves are model sequences with
 $\log g = 4.5$, black is cloud-free and the lighter curve (orange in the online version) is
cloudy with  $f_{\rm sed} = 5$.
 Small dots are a representative sample of stars taken from the 
UKIDSS Large Area Survey. Medium-sized dots are M, L and early-type T dwarfs 
with data taken from the Leggett et al. (2010, and references therein). Larger dots with error bars are the sample
presented here. Lighter datapoints  (cyan and green in the online version) represent the binaries 
WISE J1711$+$3500A and B and WISE J1217$+$1626A and B.
\label{fig6}}
\end{figure}

\begin{figure}
\includegraphics[angle=0,scale=.6]{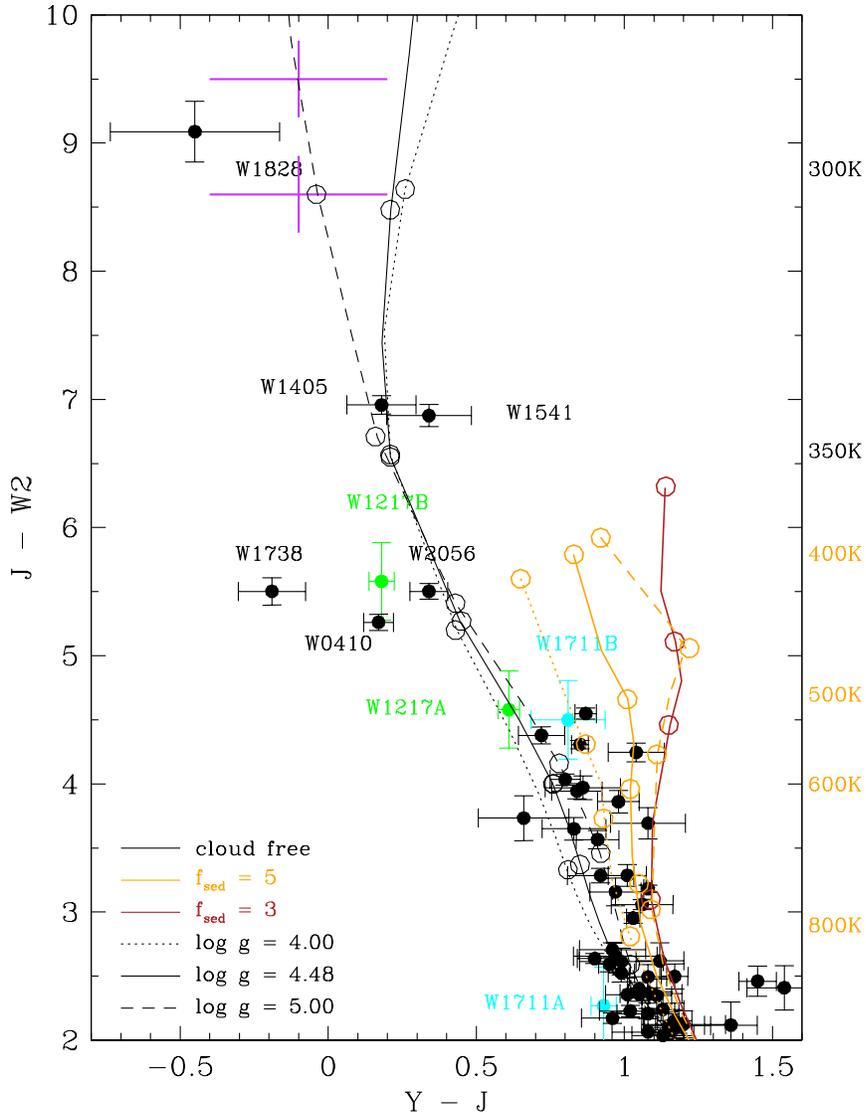}
\caption{$J -$ W2 as a function of $Y - J$. Symbols and curves are as in 
Figure 4 and 6.
\label{fig7}}
\end{figure}

\begin{figure}
\includegraphics[angle=-90,scale=.65]{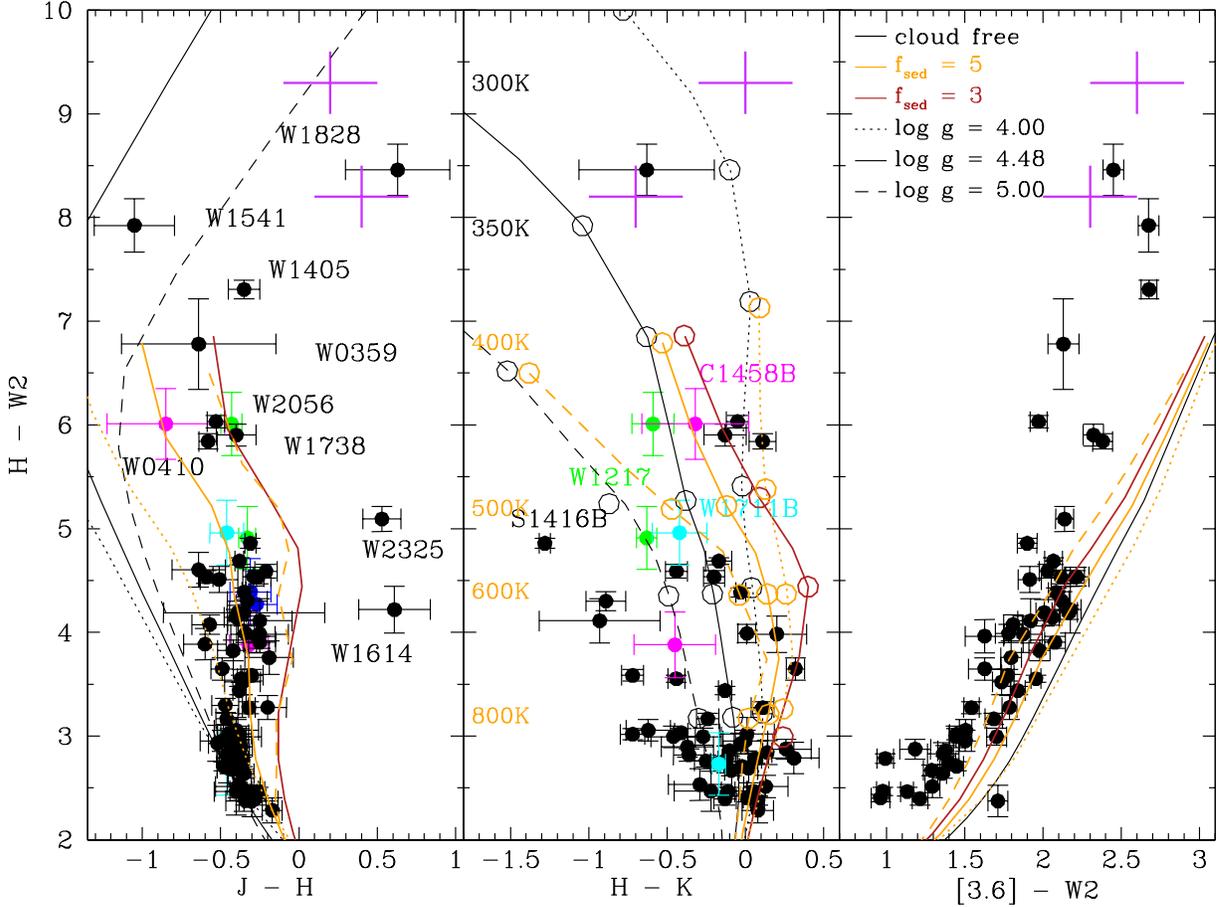}
\caption{$H -$ W2 as a function of $J - H$, $H - K$ and [3.6] $-$ W2. 
Symbols and curves are as in Figure 4 and 6.  The two T9 dwarfs WISEJ 1614$+$1739 and WISEJ 2325$-$4105
appear to be too red in $J - H$ and the Kirkpatrick et al. (2011)   photometry should be repeated
(see also Figure 3).
\label{fig8}}
\end{figure}

\begin{figure}
\includegraphics[angle=-90,scale=.6]{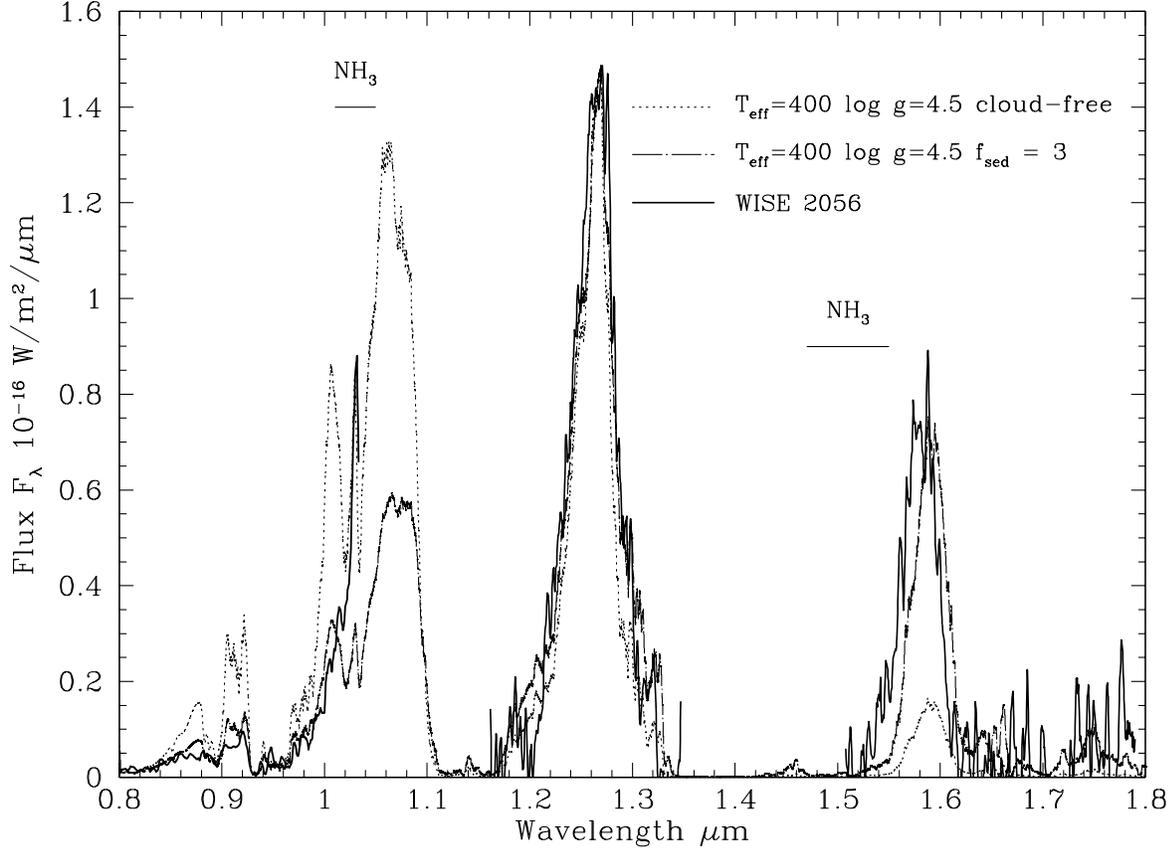}
\caption{ The near-infrared and far-red spectrum of WISEPC J2056$+$1459 from Cushing et al. (2011) and this work is shown as a black curve. Dotted and dash-dot curves are cloud-free and cloudy models with   $T_{\rm eff} = 400$~K and  $\log g = 4.5$. The models do not include vertical mixing, and so have NH$_3$ absorption at  1.03~$\mu$m and 1.52~$\mu$m that is too strong. The discrepancy at 1.6~$\mu$m is most likely due to remaining incompleteness in the CH$_4$ opacity line list. The models have been scaled to match the observations at the peak of the $J$-band.
\label{fig9}}
\end{figure}

\begin{figure}
\includegraphics[angle=-90,scale=.6]{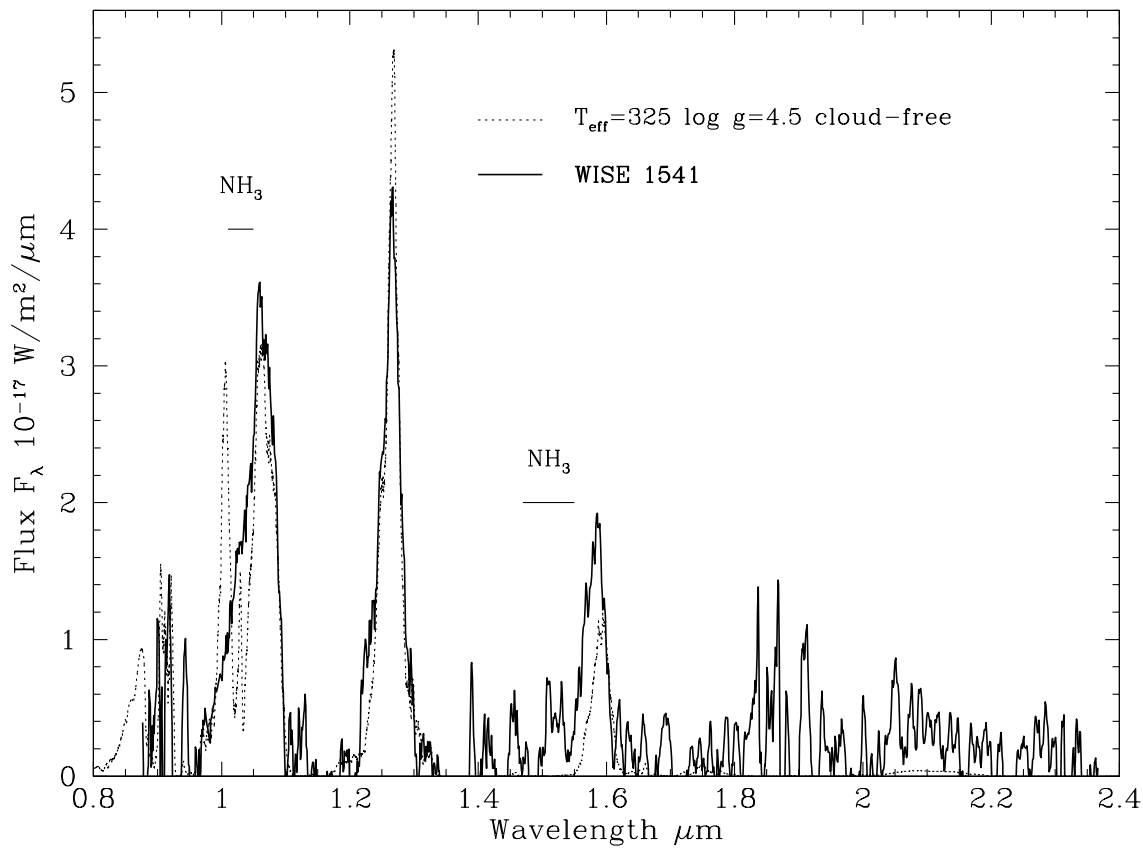}
\caption{The near-infrared spectrum of WISEPC J1541$-$2250 from Cushing et al. (2011)  is shown as a black curve. Dotted curves are cloud-free models with   $T_{\rm eff} = 325$~K and  $\log g = 4.5$. As in Figure 9, the NH$_3$ absorption features in the models are too strong due to the neglect of mixing. The model has been approximately scaled to reproduce the observed J band flux, as the observed and calculated absolute $J$ magnitude are consistent (Figure 4).
\label{fig10}}
\end{figure}





\clearpage

\begin{deluxetable}{lrrrrrrr}
\tabletypesize{\scriptsize}
\tablewidth{0pt}
\rotate
\tablecaption{NIRI $YJHK$ Photometry for {\it WISE} Y Dwarfs}
\tablehead{
\colhead{Short Name} & \colhead{$Y$(err)} & \colhead{$J$(err)} & 
\colhead{$H$(err)} & \colhead{$K$(err)} & \colhead{Exp. $Y$, $J$, $H$, 
$K$} & \colhead{Date} & Program\\
\colhead{} & \colhead{} & \colhead{} & \colhead{} & \colhead{} & 
\colhead{minutes} & \colhead{YYYYMMDD} & \colhead{} \\
}
\startdata
WISEPC J0410$+$1502 & 19.78 (0.04) & 19.44 (0.03) & 20.02 (0.05) & 19.91 (0.07) & 5, 5, 9, 9 & 
20120809/16 & GN-2012B-Q-75\\
WISEPC J1405$+$5534 & 21.41 (0.10) & 21.06 (0.06) &  21.41 (0.08) & \nodata & 9, 9, 58.5, \nodata & 
20120210 & GN-2012A-Q-106 \\
WISEPC J1541$-$2250 & 21.63 (0.13) & 21.12 (0.06)  & 22.17 (0.25) & \nodata & 18, 18, 45, \nodata & 
20120210, 20120505 & GN-2012A-Q-106 \\
WISEPC J1738$+$2732 & 20.03 (0.07) & 20.05 (0.09) & 20.45 (0.09) & 20.58 (0.10) & 5, 2.5, 9, 18 & 
20120503, 20120702 & GN-2012A-Q-106, GN-2012B-Q-75 \\
WISEPC J1828$+$2650 & 23.20 (0.17) &   23.48 (0.23) & \nodata & 23.48 (0.36) &  108, 90, \nodata, 117 & 
20120503/04/07, 20120703/07/10  & 
GN-2012A-Q-106, GN-2012B-Q-27/75 \\
WISEPC J2056$+$1459 & 19.94(0.05) & 19.43 (0.04) & 19.96 (0.04) & 20.01 (0.06) & 5, 2.5, 9, 13.5 & 
20120517, 20120609 & GN-2012A-DD-7, GN-2012A-Q-106\\
\enddata
\tablecomments{All magnitudes are Vega magnitudes, $JHK$ are on the Mauna Kea Observatories (MKO) system (Tokunaga \& Vacca 2005), $Y$ is close to the MKO system, see \S 3.}
\end{deluxetable}

\clearpage

\begin{deluxetable}{lrrrrrrrrrrr}
\tabletypesize{\scriptsize}
\tablewidth{0pt}
\rotate
\tablecaption{{\it WISE} Y Dwarfs Astrometry and Photometry}
\tablehead{
\colhead{Short Name} &  \colhead{$\pi$(err)} &  
\colhead{$Y$(err)} & \colhead{$J$(err)} & \colhead{$H$(err)} & 
\colhead{$K$(err)} &
\colhead{[3.6](err)} & \colhead{[4.5](err)} & 
\colhead{W1(err)} & \colhead{W2(err)} & \colhead{W3(err)} & \colhead{W4(err)} \\
\colhead{} & \colhead{mas}   &
\colhead{} & \colhead{}   & \colhead{} & \colhead{}    &
\colhead{} & \colhead{}  &
\colhead{} & \colhead{}   & \colhead{} & \colhead{} \\
}
\startdata
WISEPC J0410$+$1502 & 164 (24) & 19.61  (0.04) & 19.44  (0.03)   & 20.02 (0.05) & 19.91 (0.07) &  
16.56  (0.01) & 14.12 (0.01) &
$>$18.33 & 14.18 (0.06) & $>$11.86 & $>$8.90 \\
WISEPC J1405$+$5534 & 207 (39) & 21.24 (0.10) & 21.06 (0.06) &  21.41 (0.08) & \nodata & 
 16.78 (0.01) & 14.02 (0.01) &
$>$18.82 & 14.10 (0.04) & 12.43 (0.27) & $>$9.40 \\
WISEPC J1541$-$2250 & \nodata & 21.46 (0.13) & 21.12 (0.06)  & 22.17 (0.25) & \nodata & 
16.92  (0.02)  & 14.12 (0.01) &
16.74 (0.17) &	14.25 (0.06) &	$>$12.31 & $>$8.89 \\
WISEPC J1738$+$2732 & 111 (36) & 19.86 (0.08) & 20.05 (0.09) & 20.45 (0.09) & 20.58 (0.10) & 
 16.87  (0.01) &  14.42 (0.01) &
$>$18.41 & 14.55 (0.06) & 11.93 (0.19) & $>$8.98 \\
WISEPC J1828$+$2650 & 122 (13) & 23.03 (0.17) &   23.48 (0.23) & 22.85 (0.24) & 23.48 (0.36) & 
  16.84 (0.01) &   14.27 (0.01) &
$>$18.47 & 14.39 (0.06) & $>$12.53 & $>$8.75 \\
WISEPC J2056$+$1459 & \nodata & 19.77 (0.06) & 19.43 (0.04) & 19.96 (0.04) & 20.01 (0.06) & 
 15.90 (0.01)  & 13.89 (0.01) &
$>$18.25 & 13.93 (0.05) & 12.00 (0.27) & $>$8.78 \\
\enddata
\tablecomments{Parallaxes and lower limits on the {\it WISE} magnitudes are from Kirkpatrick et al. 2012. {\it WISE} data are from the All-Sky Data Release; the W1 value for WISE J1541$-$2250 is likely to be spurious, as the  {\it WISE} images show no clear W1 source at the position of the W2 source, but does show other, bluer, nearby sources.
$YJHK$ are from this work except for $H$ for WISEPC J1828$+$2650 which is from Kirkpatrick et al. 2012. The NIRI $Y$-band data have been put onto the UKIRT/WFCAM $Y$ system by subtracting 0.17 magnitudes (see \S 3).
The  IRAC magnitudes are derived by us from archived data for GO program 70062 and DDT program 551. A 30 mmag (3\%) error should be added in quadrature to the quoted random errors for the IRAC data to account for systematics. }
\end{deluxetable}

\clearpage

\begin{deluxetable}{lllrrrrrrrrrrrrrr}
\tabletypesize{\scriptsize}
\tablewidth{0pt}
\rotate
\tablecaption{Absolute Infrared Magnitudes for Morley et al. (2012) Cloudy Brown Dwarf Models}
\tablehead{
\colhead{$T_{\rm eff}$~K} & \colhead{$\log g$} & \colhead{$f_{\rm sed}$} & \colhead{$Y$} & \colhead{$J$} & \colhead{$H$} & \colhead{$K$} & \colhead{$L^{\prime}$} &  \colhead{$M$} &  
\colhead{[3.6]} & \colhead{[4.5]} &   \colhead{[5.8]} & \colhead{[8.0]} &
\colhead{W1} & \colhead{W2} & \colhead{W3} & \colhead{W4} \\
\colhead{} & \colhead{} & \colhead{} & \multicolumn{6}{c}{MKO} & \multicolumn{4}{c}{IRAC} &\multicolumn{4}{c}{\it WISE} \\

}
\startdata
400.0  &   4.00  &   2.0   &    22.72  &   21.56  &   22.11  &   21.71  &   16.77   &  14.04   &  17.99   &  14.49   &  15.85   &  15.08  &   19.12   &  14.46  &   13.47  &   11.91 \\
400.0  &   4.00   &  3.0   &    21.80    &  20.93   &  22.01  &   21.79  &   16.8   &   14.05   &  18.03   &  14.51  &   15.88   &  15.12  &   19.17  &   14.47  &   13.5   &   11.93   \\
400.0  &   4.00   &  4.0    &   21.34   &  20.63   &  22.00   &   21.86  &   16.83  &   14.07   &  18.06   &  14.53   &  15.91  &   15.14  &   19.21  &   14.49  &   13.51  &   11.93  \\
400.0  &   4.00   &  5.0   &    21.08  &   20.47  &   22.01  &   21.90   &   16.85  &   14.08   &  18.09   &  14.54   &  15.92  &   15.15  &   19.23  &   14.50   &   13.52  &   11.94  \\
400.0  &   4.48   &  2.0   &    23.40   &   21.99   &  22.06   &  22.18   &  16.82  &   14.29   &  18.06   &  14.67   &  16.18  &   15.40  &    19.23  &   14.66  &   13.79  &   12.19  \\
400.0  &   4.48   &  3.0   &    22.44  &   21.34   &  21.90   &   22.26   &  16.85   &  14.30   &   18.10   &   14.68   &  16.21  &   15.43  &   19.28   &  14.66 &  13.81   &  12.20    \\
400.0  &   4.48  &   4.0    &    21.93  &   21.02  &   21.87  &   22.32  &   16.87   &  14.31   &  18.13   &  14.69   &  16.24   &  15.45  &   19.31 &    14.68  &   13.83  &   12.21   \\
400.0  &   4.48  &   5.0    &   21.64   &  20.85   &  21.86  &   22.37   &  16.89   &  14.32   &  18.14   &  14.70    &  16.25   &  15.47   &  19.33  &   14.69  &   13.84  &   12.21  \\
\enddata
\tablecomments{The full table is available online. The full table covers $ 400 \leq T_{\rm eff}$~K $\leq 1200$, $4.00 \leq \log g \leq 5.48$ and $2.0 \leq f_{\rm sed} \leq 5.0$.}
\end{deluxetable}

\begin{deluxetable}{llrrrrrrrrrrrrrr}
\tabletypesize{\scriptsize}
\tablewidth{0pt}
\rotate
\tablecaption{Absolute Infrared Magnitudes for Saumon et al. (2012) Cloud-Free Brown Dwarf Models}
\tablehead{
\colhead{$T_{\rm eff}$~K} & \colhead{$\log g$}  & \colhead{$Y$} & \colhead{$J$} & \colhead{$H$} & \colhead{$K$} & \colhead{$L^{\prime}$} &  \colhead{$M$} &  
\colhead{[3.6]} & \colhead{[4.5]} &   \colhead{[5.8]} & \colhead{[8.0]} &
\colhead{W1} & \colhead{W2} & \colhead{W3} & \colhead{W4} \\
\colhead{} & \colhead{}  & \multicolumn{6}{c}{MKO} & \multicolumn{4}{c}{IRAC} &\multicolumn{4}{c}{\it WISE} \\
}
\startdata

200.0   &   4.00  &    34.45  &    33.46  &    33.24   &   38.37  &    22.76  &    17.72   &   25.11   &   18.17  &    20.59  &    19.15  &    26.87  &    18.18  &    17.23   &   13.78   \\
200.0   &   4.48  &    33.54  &    33.16   &   32.12  &    38.92  &    22.44  &    17.75  &    24.90   &    18.10   &    20.61  &    19.22  &    26.68  &    18.11  &    17.38   &   13.91     \\
200.0   &   5.00  &   32.76   &   33.01  &    31.21  &    39.65  &    22.02   &   17.76  &    24.55   &   17.99  &    20.57  &    19.28  &    26.38  &    18.01  &    17.55   &   14.06     \\
250.0  &    4.00  &    29.17   &   28.59  &    29.32  &    31.65  &    20.61  &    16.47  &    22.43   &   16.94  &    19.08  &    17.84 &     24.01  &    16.93  &    15.85  &    13.04     \\
250.0  &    4.48  &    28.67  &    28.39   &   28.46  &    32.33  &    20.40   &    16.52  &    22.30   &   16.89  &    19.25  &    17.97  &    23.91 &     16.89  &    16.03  &    13.15    \\ 
250.0  &    5.00  &   28.21   &   28.47  &    27.78  &    33.27  &    20.14  &    16.61  &    22.12  &    16.85  &    19.43  &    18.10  &     23.78   &   16.87  &    16.24   &   13.28     \\
300.0   &   4.00   &   25.14   &   24.92   &   26.29  &    27.04  &    18.96  &    15.46   &   20.47  &    15.95  &    17.73   &   16.71  &    21.87  &    15.92   &   14.82  &    12.55     \\
300.0   &   4.48   &   25.02  &    24.86   &   25.69  &    27.70   &    18.89  &    15.61   &   20.46  &    16.01  &    18.01    &  16.97  &    21.90   &    16.00    &   15.09  &    12.77     \\
300.0   &   5.00  &   24.86   &   24.96  &    25.13  &    28.58  &    18.69  &    15.70  &     20.32  &    15.97  &    18.25  &    17.17  &    21.82   &   15.98  &    15.32   &   12.89   
  \\
\enddata
\tablecomments{The full table is available online. The full table covers $ 200 \leq T_{\rm eff}$~K $\leq 1200$ and  $4.00 \leq \log g \leq 5.48$.}
\end{deluxetable}

\begin{deluxetable}{lrrrrrrrrrrr}
\tabletypesize{\scriptsize}
\tablewidth{0pt}
\rotate
\tablecaption{Binaries at the T/Y Dwarf Boundary}
\tablehead{
\colhead{Name} & \colhead{Spectral} & \colhead{$\pi$(err)} & 
\colhead{$Y$(err)} & \colhead{$J$(err)} & \colhead{$H$(err)} & 
\colhead{$K$(err)} & \colhead{W2(err)} & \colhead{Mass M$_{\rm Jup}$} & 
\colhead{$T_{\rm eff}$~K} & \colhead{$\log g$} & 
\colhead{Reference} \\
\colhead{} & \colhead{Type} & \colhead{mas} & 
\colhead{} & \colhead{} & \colhead{} & 
\colhead{} & \colhead{} & 
\multicolumn{3}{c}{1 Gyr/5 Gyr} & 
\colhead{} \\

}
\startdata
WISE J0458$+$6434A & T8.5 & \nodata & \nodata & 17.50 (0.07) & 17.77 
(0.11) & \nodata & 13.5 (0.3) & \nodata & \nodata & \nodata & B12 \\
WISE J0458$+$6434B & T9.5 & \nodata & \nodata & 18.48 (0.07) & 18.79 
(0.11) & \nodata & 14.4 (0.3) & \nodata & \nodata & \nodata & B12 \\
WISE J1217$+$1626A & T9 & \nodata & 18.59 (0.04) & 17.98 (0.02) & 18.31 (0.05) & 18.94 (0.04) & 13.4 (0.3) & 13/33 & 550/650 & 4.5/5.0 & L12 \\
WISE J1217$+$1626B & Y0 & \nodata & 20.26 (0.04) & 20.08 (0.03) & 20.51 (0.06) & 21.10 (0.12) & 14.5 (0.3) & 7/17 & 400/400 & 4.2/4.7 & L12 \\
CFBDS 1458$+$1013A & T9 & 34.0 (2.6) & \nodata & 19.86 (0.07 & 20.18 (0.10 ) & 20.63 (0.24) & 13.5 (0.3) &  12/35 & 550/600 & 4.5/5.0 & L12 \\
CFBDS 1458$+$1013B & Y0 & 34.0 (2.6) & \nodata & 21.66 (0.34) & 22.51 (0.16) & 22.83 (0.30) & 14.4 (0.3) & 7/17 & 350/400 & 4.1/4.6 & L12 \\
WISE J1711$+$3500A & T8  & \nodata & 18.60 (0.03) & 17.67 (0.03) & 18.13 (0.03) & 18.30 (0.03) & 15.4 (0.3) & 20/45 & 750/850 & 4.7/5.2 & L12 \\
WISE J1711$+$3500B & T9.5 & \nodata & 21.31 (0.11) & 20.50 (0.06) & 20.96 (0.09) & 21.38 (0.15) & 16.0 (0.3) & 9/23 & 450/450 & 4.3/4.8 & L12 \\
\enddata
\tablecomments{Binary discovery references are: B12 Burgasser et al. 2012, L11 Liu et al. 2011; L12 Liu et al. 2012. Trigonometric parallax for CFBDS 1458$+$1013 from Kirkpatrick et al. 2012. $YJHK$ photometry and physical properties from the discovery reference papers. Estimated resolved W2 magnitudes are from this work (\S 4.2). 
}
\end{deluxetable}

\begin{deluxetable}{llcccrr}
\tabletypesize{\scriptsize}
\tablewidth{0pt}
\tablecaption{Estimated Y0 Dwarf Properties}
\tablehead{
\colhead{Name} & \colhead{Spectral} & \colhead{$T_{\rm eff}$~K}  & \colhead{$\log g$} & \colhead{$f_{\rm sed}$} & \colhead{Mass}  & \colhead{Age} \\
\colhead{}     &  \colhead{Type}    & \colhead{}                  & \colhead{}          & \colhead{}         &
\colhead{M$_{\rm Jup}$}  & \colhead{Gyr} \\
}
\startdata
WISEPC J0410$+$1502 & Y0 & 400 -- 450 & 4.5 & 3 & 10 -- 15 & 1 -- 5 \\
WISEPC J1405$+$5534 & Y0pec? & 350 & 4.5 & cloud-free & 10 -- 15 & 5 -- 10 \\
WISEPC J1541$-$2250 & Y0.5 & 300 -- 350 & 4.0 -- 4.5 & cloud-free & 5 -- 13 & 1 -- 10 \\
WISEPC J1738$+$2732 & Y0 & 400 -- 450 & 4.5 & 3 & 10 -- 15 & 1 -- 5 \\
WISEPC J2056$+$1459 & Y0 & 400 -- 450 & 4.5 & 3 & 10 -- 15 & 1 -- 5 \\
\enddata
\tablecomments{Spectral types are from Kirkpatrick et al. 2012.}
\end{deluxetable}

\clearpage

\begin{deluxetable}{lrrrrrrrrrrr}
\tabletypesize{\scriptsize}
\tablewidth{0pt}
\tablecaption{WISEPC J1828$+$2650 as a Binary}
\tablehead{
\colhead{Component} & \colhead{Y} & \colhead{J} & \colhead{H} & 
\colhead{K} & \colhead{[3.6]}  & \colhead{[4.5]} & \colhead{W2}   & \colhead{$T_{\rm eff}$~K}  & \colhead{$\log g$}  & \colhead{Mass}  & \colhead{Age} \\
\colhead{} & \colhead{} & \colhead{} & \colhead{} & 
\colhead{} & \colhead{}  & \colhead{} & \colhead{}   & \colhead{}  & \colhead{}  & \colhead{M$_{\rm Jup}$}  & \colhead{Gyr} \\
}
\startdata
AB & 23.03  & 23.48 & 22.85 & 23.48 & 16.84  & 14.27 & 14.39 & & & & \\
A &     23.6      &        23.7  &       23.3   &        24.0  &        17.4  &       15.0     &     15.1 & 325 & 4.5 & 10 & 2  \\                                                        
B &     24.6      &        24.7  &       24.5   &        24.5  &        17.8  &       15.1   &     15.2  & 300 & 4.0 & 7 & 2\\
\enddata
\end{deluxetable}

\clearpage

\begin{deluxetable}{llrrcrrrr}
\tabletypesize{\scriptsize}
\tablewidth{0pt}
\tablecaption{Appendix Data Table Sample}
\tablehead{
\colhead{Name} & \colhead{Other} & \colhead{RA} & \colhead{Decl.} & 
\colhead{Spectral} & \colhead{Dist. Mod.}  & \colhead{M$_J$} & \colhead{i}  & \colhead{z} \\
\colhead{} & \colhead{Name} & \colhead{HHMMSS.SS} & \colhead{SDDMMSS.S} & 
\colhead{Type} & \colhead{$M - m$}  & \colhead{} & \colhead{} & \colhead{} \\
}
\startdata
ULAS0034-0052 & & 00 34 02.77 & $-$00 52 06.7 & T8.5 & $-$0.82 & 17.67 & & 22.00 \\
2MASS0034+0523 & & 00 34 51.57 & 5 23 05.0 & T6.5 & 0.11 & 15.22 & 24.47 & 18.93\\
2MASS0050-3322 & & 00 50 19.94 & $-$33 22 40.2 & T7 & $-$0.12 & 15.53 & & \\
CFBDS0059-0114 & & 00 59 10.9 & $-$01 14 01.3 & T8.5 & 0.07 & 18.13 & & 21.73\\
WISEP0148-7202 & & 01 48 07.25 & $-$72 02 58.7 & T9.5 & & & & \\
\enddata
\tablecomments{The full table is available online.}
\end{deluxetable}

\end{document}